%% file: new0007043.tex
\documentclass[a4,12pt]{article}
\usepackage{epsfig}
\begin{document}


\title{Inflaton Particles in Reheating}
\author{
A.~B.~Henriques$^a$ and
R.~G.~Moorhouse$^b$\\[.4cm]
\small $^a$Departamento de Fisica/CENTRA,Instituto Superior Tecnico,
1096 Lisbon, Portugal.\\
\small $^b$Department of Physics and Astronomy,University of Glasgow,
Glasgow G12 8QQ, U.K.\\
}

\maketitle

\begin{abstract}
\noindent
 In many theories of reheating starting from the classical spatially homogeneous 
 inflaton field, its accompanying inhomogeneous part (which arises from
 primordial quantum fluctuations) is treated as a first order perturbation.
 We examine some consequences of treating it non-perturbatively in a 
 model where a first order treatment is invalid. In particular we consider
 effects on the long-wavelength curvature parameter $\zeta$.
\end{abstract} 
\section{Introduction}
Starting from minimal models of inflation with one scalar field, $\varphi$, 
the addition of a scalar field, $\chi$, interacting with it has been much
investigated, particularly in the context of parametric resonance
 \cite{TRA,KLS1,etc,KLS2,K,KT,Boy,Fin}. The inflaton field, $\varphi$, is 
 equal to $\varphi_0+\varphi_1$, where $\varphi_0$ is the classical field 
giving rise to inflation through a potential $V(\varphi)$ and $\varphi_1$
arises as a primordial quantum perturbation. $\varphi_1$ is associated 
with perturbation to the metric also arising from primordial quantum 
perturbations. In longitudinal gauge (and in the usual case of absence
of space-space off-diagonal elements of the stress-energy tensor) 
  the FRW metric appears with the perturbing field $\psi$ as
\begin{equation}\label{eq1}
ds^2 = a(\tau)^2(1+2\psi)d\tau^2-a(\tau)^2(1-2\psi)\delta_{ij}dx^idx^j. 
\end{equation}
The equations of motion imply that $\varphi_1$ and $\psi$ arise having
identical $k$-component quantum creation and annihilation operators. In
this minimal model the small $k$ (long wavelength) components of
 $\varphi_1$ and $\psi$ are responsible for the observed cosmic 
microwave background radiation fluctuations (CMBRF). Though an additional
 field $\chi$ could also have an influential classical part, thus forming 
a two-field inflation model we shall consider single field inflation 
with $\chi$ only arising from its own primordial quantum fluctuations.
In fact with the well-used interaction $g^2\varphi^2\chi^2$ 
\cite{TRA,KLS1,etc,KLS2,K} there are 
reasons for any $\chi$ field being strongly suppressed before reheating 
begins. So in this paper on reheating we have initially three quantum
 fields $\varphi_1, \psi$ and $\chi$ ,the latter being composed of 
different quantum operator components to those of $\varphi_1$ and 
$\psi$, which are identical. (The $\chi$ field has a seemingly negligable 
 value at the beginning of reheating compared with the other two but can
 increase to logarithmically comparable values through resonance effects,
 as we shall see.)

These quantum fields decohere into classical stochastic fields. There 
are a number of treatments of decoherence but the work of Polarski 
 and Starobinsky \cite{PS}, giving a gradual decoherence with the expansion 
 of the universe and the multiplication in the number of particles, is
 particularly relevant. The stochastic fields inherit, in transmuted 
form, important properties of their quantum progenitors; in particular 
ensemble averaging of the products of stochastic fields gives similar 
results to taking vacuum expectation values of corresponding products 
 of quantum fields.

For example, arising from the interaction $g^2\varphi^2\chi^2$, the 
equation of motion of the $\varphi_0$ field contains a term 
$g^2\varphi_0\chi^2$ and the only simple way to deal with this is to 
take an ensemble average of the $\chi^2$ term, which we denote by
$<\chi^2>$. In terms of the mode functions this gives exactly 
the same expression as the vacuum expection value, $<\chi^2>_0$, 
of the corresponding quantum fields. Thus, since the mode functions
 themselves are smooth, there is a smooth transition from the quantum 
 era to the classical stochastic era.

 A very significant use we make of ensemble averaging is in the 
 calculation of the Hubble parameter, $H (a'/a=aH)$, The picture 
 during the inflation period is that the classical inflaton field
 dominates the pressure and energy densities and gives $H$. However
 it can happen during reheating - and does in the examples we shall 
 treat - that with the diminution of $\varphi_0$ the contribution of
 $\varphi_1$ to the energy density rapidly becomes larger than that 
 of $\varphi_0$, and its contribution to the spatially homogeneous 
 Einstein equations and the calculation of $H$ cannot be ignored or 
 treated as a perturbation. Thus if $G^{\mu 0}_{\nu}$ is the 
 Einstein tensor evaluated from the unperturbed ($\psi=0$) metric,
$ds^2 = a(\tau)^2d\tau^2-a(\tau)^2\delta_{ij}dx^idx^j$, we write  
 \begin{equation}\label{eq2}
. G^{\mu 0}_{\nu} = <T^{\mu}_{\nu}>.
\end{equation}
 As one example this means that a term $(\varphi'_0)^2$ appearing on 
 the RHS of Eq.(\ref{eq2}) receives an addition
 $<(\varphi'_1)^2> + <(\chi'_1)^2>$. Thus the homogeneous energy and 
 pressure densities, $\rho_0$ and $p_0$, during reheating and 
 consequently the magnitude of $H$, may be largely governed by such
additions.

As an illustration of some results of this approach in our model 
 we evaluate the well-known parameter $\zeta$ \cite{BST,LYTH} which, 
 as a function of the wave number $k$, is defined \cite{MUK} through 
 the $k$-component, $\psi_k$,of the metric perturbation $\psi$
 \begin{equation}\label{eq3}
\zeta_k = \frac{2}{3}(H^{-1} \dot\psi_k + \psi_k)/(1+w) + \psi_k  
\end{equation}
For those small values of $k$ corresponding to the observed CMBRF we
 have that $k^2/a^2 \ll H^2$ for any $H$ in the reheating period and
 thus $\zeta_k \approx  -{\cal R}_{k}$ where $ {\cal R}_{k}$ is the
 curvature perturbation\cite{LLbook}. For such $k$ and with the assumption 
 of adiabaticity $\zeta$ is constant through reheat and to the matter
 era \cite{BST,LYTH,MUK,LLbook,LL,WMLL,LLMW}. 
 So it has often been used as a tool in calculating the CMBRF 
 in various models of inflation; there has been considerable interest
 in the effect of parametric resonance on this issue 
\cite{LLMW,BASS,BGMK}. We shall examine the variation of 
 $\zeta$ both analytically and numerically in the context outlined
 above.

 \section{Decay of the Inflaton and Preheating}
 In this paper we shall not treat true reheating (otherwise defrosting) 
 in which there is conversion in large part to a thermal (relativistic)
 fluid. Rather we shall deal with a first stage of the conversion
of the classical field $\varphi_0$ into particles of the fields
 $\chi$ and $\varphi_1$. This, specially with respect to $\chi$,
 has been named preheating and we shall see that the particles 
 $\varphi_1$ can stand equally importantly in this respect. This being 
 understood we shall continue to use the word reheating to include
 also this preheating stage.
 \subsection{The equations of motion} 

 The equations of motion divide into two classes, the spatially 
 homogeneous and the spatially non-homogeneous. The former are 
 those for those two variables which are functions of time only,
 $a(\tau)$ and $\varphi_0(\tau)$. The latter are those in the 
 scalar fields $\varphi_1({\bf x},\tau), \chi({\bf x},\tau), 
 \psi({\bf x},\tau)$ which are functions of space and time; these
 equations we shall express in terms of the Fourier component 
 mode functions $\varphi_{k}(\tau), \chi_k(\tau), \psi_k(\tau)$ 
 in a way we shall describe below.

 We emphasize that this is not a division into non-perturbed 
 and perturbation equations. The equations are to maximum order 
 in $\varphi_1$ and $\chi$ as well as $\varphi_0$. We do treat 
 $\psi$ as a perturbation and we examine the validity of this 
 approximation.

 We now specify the model we use more precisely \cite{HMoor}.

The Lagrangian is
 \begin{equation}\label{eq4}
L=\int d^4x\sqrt{-g}\lbrack\frac{1}{2}\varphi^{,\alpha}\varphi_{,\alpha}+
\frac{1}{2}\chi^{,\alpha}\chi_{,\alpha}-V(\varphi)-V(\chi)-
V_{int}(\varphi,\chi)\rbrack 
\end{equation}
 where $\varphi=\varphi_0 + \varphi_1$ and the longitudinal gauge 
 metric is given by Eq.(\ref{eq1}). This leads to the energy-momentum 
 tensor
 \begin{equation}\label{eq5}
T^{\mu}_{\nu} =\varphi^{,\mu}\varphi_{,\nu}+\chi^{,\mu}\chi_{,\nu}-
\lbrack\frac{1}{2}\varphi^{,\alpha}\varphi_{,\alpha}+
\frac{1}{2}\chi^{,\alpha}\chi_{,\alpha}-
V(\varphi)-V(\chi)-V_{int}\rbrack \delta^{\mu}_{\nu}
\end{equation}  
We take the field potentials during reheating to be
\begin{equation}\label{eq6}
 V(\varphi)=\frac{1}{2}m^2\varphi^2 ; V(\chi)=\frac{1}{2}M^2\chi^2; 
  V_{int}(\varphi,\chi) = \frac{1}{2} g^2 \varphi^2 \chi^2. 
\end{equation} 
 The potential $V(\varphi)=\frac{1}{2}m^2\varphi^2$ is that used in 
 chaotic inflation theory 
 but is applied in this paper only in the reheating period; we use a 
 different, but smoothly joining, potential in the inflationary era 
 \cite{HMoor}.

The classical stochastic fields are of the form

  \begin{equation}\label{eq7}  \varphi_1({\bf x},\tau) = 
 \int \frac{d^{3}k}{(2\pi)^{\frac{3}{2}}} 
\lbrack e({\bf k})\varphi_{\bf k}(\tau)\exp(i{\bf k.x}) 
+  e^*({\bf k})\varphi_{\bf k}^*(\tau)\exp(-i{\bf k.x}) \rbrack, 
\end{equation}
  \begin{equation}\label{eq8}  \chi({\bf x},\tau) = 
 \int \frac{d^{3}k}{(2\pi)^{\frac{3}{2}}} 
\lbrack d(\bf k)\chi_{\bf k}(\tau)\exp(i{\bf k.x}) 
+  d^*(\bf k)\chi_{\bf k}^*(\tau)\exp(-i{\bf k.x}) \rbrack,
\end{equation}

 where $e(\bf k)$ and $d(\bf k)$ are time-independent separately 
 $\delta$-correlated Gaussian variables \cite{PS} such that, where 
 $\langle .... \rangle$ denotes the average,
 \begin{equation}\label{eq9}
 \langle e({\bf k})e^*({\bf k}') \rangle=
 \langle d({\bf k})d^*({\bf k}') \rangle
 =\frac{1}{2} \delta^3({\bf k -k'});
 \end{equation}  
 with all other two-variable averages being zero. 
 $\varphi_{\bf k}(\tau)$ and $\chi_{\bf k}(\tau)$ are the mode 
 functions in which we express the equations of motion. The field
 $\psi({\bf x},\tau)$ is expressed as Eq.(\ref{eq7}) with the 
 same gaussian variables but the mode functions being
 $\psi_{\bf k}(\tau)$.

 The properties of the gaussian variables given above result,
 for example, in the ensemble average
   \begin{equation}\label{eq10}
 \langle\varphi_1(x,\tau)^2\rangle=(2\pi)^{-3}\int d^3k 
\varphi_{\bf k}\varphi_{\bf k}^*
\end{equation}

 (The originating quantum fields are expressed as Eqs.(\ref{eq7},\ref{eq8})
 except that the gaussian variables are replaced by appropriate 
 creation and annihilation operators; the vacuum expection values
 $<....>_0$ of products of quantum fields are unchanged in form from 
 the ensemble averages $<....>$; so ,for example,   
 $\langle\varphi_1(x,\tau)^2\rangle_0$ repeats the same form as 
 Eq.(\ref{eq10}).)

Integrals such as that in Eq.(\ref{eq10}) occur throughout the equations of 
motion and must be evaluated numerically \footnote{This is discussed in 
 detail in ref.\cite{HMoor}.}and we have to adopt
 a finite range of wave number, $k$. If the integrals diverge as 
$k \rightarrow \infty$ then the upper limit of the $k$ integration forms a
cut-off which is the crudest way of dealing with such ultra-violet 
divergences. Two points of view may be taken on this. Firstly this may be 
considered equivalent to a renormalization procedure in mass and other 
quantities. Secondly the Lagrangian used may be considered as an effective 
Lagrangian which has absorbed extra degrees of freedom coming from 
supersymmetry which eliminate divergences at higher momentum, the cut-off 
representing this effect. In our numerical work we have generally taken a 
cut-off to correspond to a wavelength of $H^{-1}$ so that 
$k_{cutoff} \approx 2\pi aH$; this has been used in other work
involving ensemble averages. We discuss further on the effect of different 
evaluations.

We can now proceed to write down the equations of motion. As the 
 above remarks explain these are the same in the quantum regime as in
 the classical stochastic regime. First we must consider 
 $T^{\mu}_{\nu}$.
 Defining the density and pressure homogeneous parts of the 
 energy-momentum tensor as $\rho_h(\tau) \equiv \langle T^0_0 \rangle$ 
 and $p_h(\tau)\delta^i_j \equiv -\langle T^i_j \rangle$ we find, after 
 ensemble averaging, that
\begin{equation}\label{eq11}
 \rho_h(\tau) = \frac{1}{2 a^2} \lbrack \eta 
+a^2\bar{m}^2(\varphi_0^2+\langle \varphi_1^2 \rangle)
+a^2M^2 \langle \chi^2 \rangle
+\langle \varphi_{1,i}^2 \rangle+\langle \chi_{,i}^2 \rangle
 -4\varphi'_0\langle\varphi_1'\psi\rangle\rbrack ,
\end{equation} 
\begin{equation}\label{eq12}
 p_h(\tau)=\frac{1}{2 a^2} \lbrack \eta
-a^2\bar{m}^2(\varphi_0^2+\langle \varphi_1^2 \rangle)
-a^2M^2 \langle \chi^2 \rangle
-\langle \varphi_{1,i}^2 \rangle/3-\langle \chi_{,i}^2 \rangle/3
 -4\varphi'_0\langle\varphi_1'\psi\rangle\rbrack , 
\end{equation} 
\begin{equation}\label{eq13}
 \eta =  
 \varphi_0'^2+\langle\varphi_1'^2\rangle+\langle \chi'^2 \rangle
\end{equation} 
\begin{equation}\label{eq14}
\bar{m}^2 \equiv m^2+g^2\langle \chi^2 \rangle.
\end{equation}
and we have the Friedmann equation
\begin{equation}\label{eq15}
(a'/a)^2= \frac{8\pi G}{3}a^2\rho_h(\tau).
\end{equation}

As in Eq.(\ref{eq10}) the averages are independent of ${\bf x}$ ensuring 
 the spatially homogeneity of the R.H.S. of Eq.(\ref{eq15}). We also see
 from Eq.(\ref{eq10}) that Eq.(\ref{eq15}) is equally valid in the quantum 
 regime since the formalism ensures that the vev of the quantum operators 
 equals the average over the stochastic variables.

The other spatially homogeneous equation is 
\begin{equation}\label{eq16}
\varphi_0''+2(a'/a)\varphi_0'+a^2 \bar {m}^2 \varphi_0
-4\langle \psi'\varphi_1' \rangle-4\langle \psi\nabla^2\varphi_1\rangle
+2a^2\bar{m}^2 \langle \psi\varphi_1 \rangle= 0
\end{equation}

We note the combined reaction of the inhomogeneous fields $\varphi_1$ and 
 $\psi$ on the homogeneous inflaton field through the ensemble averaging,
 similarly to that in Eqs. (\ref{eq11}) and (\ref{eq12}).

There remain the 3 spatially non-homogeneous equations which we write in 
the $k$-component form. These components are specified as the complex mode 
functions $\chi_k$, $\varphi_k$ and $\psi_k$. Their wave number dependence, 
given  by the succeeding equations, is only on 
 $k \equiv \left|{\bf k}\right|$ .
 
\begin{equation}\label{eq17}
\chi_k''+2(a'/a) \chi_k'+(k^2+a^2\bar{M}^2)\chi_k = 0 
\end{equation}
where $\bar{M}^2$ is a function of $\tau$ given by 
 \begin{equation}\label{eq18}
\bar{M}^2(\tau)=
M^2+g^2\varphi_0(\tau)^2+g^2\langle\varphi_1({\bf x},\tau)^2\rangle_0
\end{equation}

\begin{equation}\label{eq19}
\varphi_k''+2(a'/a)\varphi_k'+(k^2+a^2 \bar {m}^2) \varphi_k -
4\varphi_0'\psi_{k}'+
2a^2\bar{m}^2\varphi_0\psi_k= 0
\end{equation}                  
Though not indicated these equations actually hold for each separate value of
 ${\bf k}$; thus consistency of Eq.(\ref{eq19}) justifies our previous 
 statements that $\psi_{\bf k}$, 
 the mode function of the metric perturbation, should be associated with the 
 the same gaussian operators (or,in the quantum regime, with the same 
 quantum operators) as $\varphi_{\bf k}$:
 \begin{equation}\label{eq20}  \psi({\bf x},\tau) = 
 \int \frac{d^{3}k}{(2\pi)^{\frac{3}{2}}} 
\lbrack e({\bf k})\psi_{\bf k}(\tau)\exp(i{\bf k.x}) 
+  e^*({\bf k})\psi_{\bf k}^*(\tau)\exp(-i{\bf k.x}) \rbrack 
\end{equation}          
where since $\psi({\bf x},\tau)$ is dimensionless $\psi_{\bf k}$ has 
 dimension $(mass)^{-\frac{3}{2}}$. This associates the metric perturbation 
 with the inhomogeneous part of the inflaton field without assigning 
 priority to  either. But the stochastic variables of $\chi_{\bf k}$ are 
 independent. Thus no terms in $\psi_{\bf k}$ appear in Eq.(\ref{eq17});
 they are forbidden through ensemble averaging.. 

The mode equation for $\psi$ is

\begin{equation}\label{eq21}
\psi_k''+3\psi_k'(a'/a) +\psi_k(2(a'/a)'+(a'/a)^2)= 4\pi G a^2 \delta p_k
\end{equation}
 where $\delta p_k(\delta \rho_k)$ is the $k$-component of the
non-homogeneous part of the momentum (energy):  

\begin{equation}\label{eq22}
 \delta p_k = \frac{1}{a^2}\lbrack 
-(\eta+\langle\varphi_{1,i}\varphi_{1,i}\rangle/3
+\langle\chi_{,i}\chi_{,i}\rangle/3 )\psi_k 
+ \varphi_0'\varphi_k'
-2\varphi_k'\langle \psi\varphi_1' \rangle
 -a^2\bar{m}^2\varphi_0\varphi_k \rbrack.
\end{equation}
In addition there is the time-space Einstein equation which acts as 
an equation of constraint on the initial values. (We use it to fix the 
value of $\psi'$ at the beginning of reheating.)

\begin{equation}\label{eq23}
 \psi'_k + (a'/a)\psi_k = 4\pi G\varphi_0' \varphi_k
 \end{equation}

\subsection{Initial conditions}

We need to have the values of $H$, the fields and their 
 time-derivatives at the end ofinflation to supply the initial 
conditions for the reheating equations of motion of the previous
 section, 2.1. We have chosen to use a specific 
 inflationary model having the advantage that the solutions are 
 analytically expressible. This is power-law inflation with an exponential 
 potential: $V =  Uexp(-\lambda\varphi)$ where $U,\lambda$ are constants.
 This potential of the inflation era stands in place of the potential
 $V(\varphi)=\frac{1}{2}m^2\varphi^2$ of the
 reheating era but otherwise the Lagrangians are the same
 and in particular both have the interaction potential 
 $V_{int}(\varphi,\chi) = \frac{1}{2} g^2 \varphi^2 \chi^2$.
 We have ensured the correct continuity by imposing the Lichnerowicz 
 conditions \cite{DER}, as well as using the equations of motion 
 and constraint appropriate to each era at the boundary \cite{HMoor}.   
 
 An essential feature in maintaining the analytic form of power-law 
 inflation is that the $\chi$-field be negligable through the relevant 
 era. This is ensured by the field interaction term in the 
 Lagrangian. A simplified version of the mode equation, 
 Eq.(\ref{eq17}), in the inflationary era is 
 \begin{equation}\label{eq24}
(a\chi_k)''+[k^2+a^2 M^2+g^2 a^2 \varphi_0^2-a''/a](a\chi_k)=0
 \end{equation}
and $\varphi_0^2$ is greater than, or of the order of, $m_{Pl}^2$ 
through most of the inflationary era. Thus the term in square brackets
 is large and positive resulting in a quasi-periodic type solution 
 for $a\chi_k$.
 The many efold increase of $a$ during the inflationary era indicates 
an exceedingly small value for $\chi_k$ at the beginning of reheat 
\cite{BassV3,JS}, which may be as little as $10^{-50}$ of its 
 initial value. This is an important qualitative feature of our
 initial conditions. We have taken the initial value of $\chi$ to be 
 $10^{-n}m_{Planck}^{-1/2}$ where $n$ is of the order of 30. 
 (For such small mode functions in the beginning of reheat the transition 
 to classical stochastic functions cannot yet be made and 
 the quantum complex formalism should be retained.)

 For power-law inflation the scale factor $a \propto (\tau_i - \tau)^p$ 
 and we are free to choose $p$ by specifying $\lambda$ in the 
 potential \cite{HMoor}. In the numerical results quoted we have chosen 
 $p=-1.1$ ($p--1$ corresponds to exponential inflation).

 \subsection{The metric perturbation,$\psi$}

As noted above $\psi$ is treated perturbatively in the equations of
motion and we check the validity of this, in each particular case 
used, in the following sense. We shall discuss $\psi$ as having developed into 
a classical stochastic field, Eq.(\ref{eq20}). The indeterminancy  represented 
 by the variables $e$ forces us to consider the average over the product 
of these variables so that we evaluate $<(\psi(x,\tau)^2>$ the result being
$ \langle\psi(x,\tau)^2\rangle=
(2\pi)^{-3}\int d^3k \psi_k(\tau)\psi_k(\tau)^*$
the same for every value of $x$. We require that $\sqrt{<(\psi(x,\tau)^2>}$    
be small compared with unity, since the relevant metric coefficient
 is $a(\tau)^2(1+2\psi)$. Our viewpoint is that in any 
particular  case satisfaction of this requirement 
forms sufficient justification for the perturbative
 approach, because the only way we can mount a comparison of the revised 
 metric coefficient with $a(\tau)^2$ is when we consider the basic equations 
 to be those in configuration space, and then indeed the revision is just a 
 perturbation through all space-time\cite{HMoor}. In the cases shown in the
 figures $\sqrt{<(\psi(x,\tau)^2>}$ is less than $10^{-2.5}$ throughout
 the range shown.

 \section{Variation of $\zeta$ During Reheating}
 
Using the equations of motion of the preceeding section we can now 
analyse the behaviour of $\zeta$ to see under what conditions it 
may be constant and what may cause it to vary.

Multiplying Eq.(\ref{eq3}) by $\frac{3}{2}H(1+w)$ and differentiating 
with respect to cosmic time we find after some manipulation and using
Eqs.(\ref{eq11}) and (\ref{eq12}) 

\begin{equation}\label{eq25}
\frac{3}{2}H(1+w)\dot \zeta=\ddot{\psi}+H\dot \psi+2\dot H \psi-
(\dot \psi+H\psi)\frac{d}{dt}\ln(\rho_h+p_h)
\end{equation}

\begin{equation}\label{eq26}
\frac{d}{dt}\ln(\rho_h+p_h) =
2\frac{\dot{\varphi}_0\ddot{\varphi}_0+
\langle\dot{\varphi}_1\ddot{\varphi}_1\rangle+
\langle\dot{\chi}\ddot{\chi}\rangle
+\langle\dot{\varphi}_{1,i}\ddot{\varphi}_{1,i}\rangle/3+
\langle\dot{\chi}_{,i}\ddot{\chi}_{,i}\rangle/3 }
{\dot{\varphi}_0^2+\langle\dot{\varphi}_1^2\rangle+\langle\dot{\chi}^2\rangle
+\langle\dot{\varphi}_{1,i}\dot{\varphi}_{1,i}\rangle/3+
\langle\dot{\chi}_{,i}\dot{\chi}_{,i}\rangle/3 }
\end{equation}
where we have omitted the last terms in Eqs.(\ref{eq11}) and (\ref{eq12})
as giving rise to terms of second order in $\psi$. (Indeed including 
them would make no difference to the arguments we shall give.)

 We now compare the RHS of this equation with an equation for 
 $\psi$ deduced from the Einstein equations. Besides 
 Eq.(\ref{eq21}) there is also the time-time equation in $\psi$:
\begin{equation}\label{eq27}
 \nabla^2 \psi_k-3(\psi_k' + (a'/a)\psi_k)(a'/a)= 
4\pi G a^2 \delta \rho_k.
\end{equation}
 Our concern is with $\psi_k$ of wave numbers relevant to the 
 CMBRF, so we can drop the first term in this equation. Then
 subtracting it from Eq.(\ref{eq21}) and using cosmic time we 
 obtain
\begin{equation}\label{eq28}
\ddot{\psi}+7H\dot \psi+2(\dot H + 3H^2) \psi=
-4\pi G(2\bar{m}^2\varphi_0\varphi_k+\xi \psi_k)
\end{equation}

\begin{equation}\label{eq29}
\xi \equiv 
\frac{4}{3a^2}(\langle \varphi_{1,i}^2 \rangle)+\langle \chi_{,i}^2 \rangle)
\end{equation}

Now eliminate $\varphi_0$ by Eq.(\ref{eq16}) and $\varphi_0'$ by
Eq.(\ref{eq23}) to get

\begin{equation}\label{eq30}
\ddot{\psi}_k+(H-2\ddot{\varphi}_0/\dot{\varphi}_0+
 8\langle\dot{\psi}\dot{\varphi}_1 \rangle/\dot{\varphi}_0)\dot{\psi}_k+
(2\dot H-2H\ddot{\varphi}_0/\dot{\varphi}_0+
 8H\langle\dot{\psi}\dot{\varphi}_1 \rangle/\dot{\varphi}_0+
4\pi G \xi)\psi_k =0
\end{equation}

Eq.(\ref{eq30}) has some structure in common with Eq.(\ref{eq25}). Thus
putting all second order terms in the stochastic fields equal to zero
both the RHS of Eq.(\ref{eq25}) and the LHS of Eq.(\ref{eq30}) reduce to 
the expression
$\ddot{\psi}_k+(H-2\ddot{\varphi}_0/\dot{\varphi}_0)\dot{\psi}_k+
(2\dot H-2H\ddot{\varphi}_0/\dot{\varphi}_0)\psi_k$, implying that
$\dot{\zeta}=0$. This essentially replicates the conditions under 
which Mukhanov et al.\cite{MUK} demonstrate the constancy of $\zeta$. 
Under our extended equations it seems most unlikely that $\zeta$ be 
constant except when $\varphi_0$ dominates and thus the quadratic 
terms, $<...>$, can be neglected; in our model this is so 
at the beginning of reheat but then $\varphi_0$ rapidly decreases
and so we should expect $\zeta$ to be no longer constant. Our 
numerical calculations do indeed show this expected behaviour 
of $\zeta$.

It is interesting that in our theory, where $\varphi_1$ is 
 not treated perturbatively, then $\zeta$ is still not apparently 
constant when we eliminate the $\chi$ field so that $\chi = 0$ in
 Eq.(\ref{eq25}) and Eq.(\ref{eq30}), as is born out
numerically. In this connection we note that one of the relevant 
 ways of distinguishing entropic from non-entropic reheating is to
 evaluate $\frac{\delta\rho}{\dot \rho}-\frac{\delta p}{\dot p}$;
 for the reheating to be non-entropic this should be zero \cite{LLMW}. 
 The expression we find in our formalism and with $\varphi_1$ not 
 treated perturbatively gives
 $\frac{\delta\rho}{\dot \rho}-\frac{\delta p}{\dot p} \ne 0$
 even when $\chi$ is not present in the theory.

 In the figures we have illustrated some examples from our numerical
 calculations.
 \footnote{The figures in the previous versions of this paper were 
 wrong; the numerical code for the solution of the equations had
 an elementary error. Some details of the coding used here are 
 given in the authors' paper hep-ph/0109218.} 
 Our interest is in the qualitative features of reheat 
 in a model compatible with known observations rather than in making 
 precise comparisons with data. So we are interested in values of the 
 parameters  that roughly supply  the usual requirements of inflation 
 and the magnitude of the CMBR fluctuations. The results we give are 
 for the values $m = 10^{-7}m_{Pl}$, initial $\varphi_0 = 0.3$,
  $p = -1.1, M/m=0.02$ and $g/m=2\times 10^4$.Throughout we quote 
 dimensionful results and parameters in units such that 
 $\hbar = c = G =1$. 

 In Fig.1 we show the energy densities of $\varphi_0$ and $\varphi_1$;
we see that in this model the energy density of $\varphi_1$ soon becomes
 comparable to that of $\varphi_0$ and finally somewhat exceeds it;  it is 
 only at the beginning of preheating that $\varphi_1$ can be treated as
a perturbation of $\varphi_0$. Fig.2 shows how $\zeta$ is initially constant, 
 that is when $\varphi_1$ is much less than $\varphi_0$, but then develops 
 strong fluctuations and later increases. Fig.3 shows $\zeta$ developing
 resonance at 5 efolds after the end of the inflation era but we note that 
 $\zeta$, as shown in Fig.2, has begun to fluctuate strongly while $\chi$ 
 is still negligable thus showing the influence of quadratic terms in $\varphi$.
 And in Fig.4 we give the $\zeta$ that results when we have eliminated 
 $\chi$ from the  equations, again showing the influence of $\varphi_1$ 
 and $\varphi_0$ quadratic terms in the fluctuations of $\zeta$.

\section{Summary and Discussion}

 We have considered a single field ($\varphi$) model of inflation, with
 another scalar field ($\chi$) which is naturally quiescent during 
 inflation and becomes active post inflation with the possibility 
 of parametric resonance and preheating as the classical part,
 $\varphi_0$, of the inflaton field decreases and oscillates. We 
 adopted a method having stochastic variables naturally succeeding the 
 quantum operators of the early inflation era and have developed 
 equations for the post-inflation period by taking averages over the 
 stochastic bilinear forms in the scalar fields. This enables a 
 non-perturbative treatment of the stochastic scalar fields 
 $\chi$ and $\varphi_1$ (where $\varphi = \varphi_0 +\varphi_1$) 
 though the metric perturbation field, $\psi$, has to be treated 
 perturbatively.

In our results as $\varphi_0$ decreases, the energy density of the $\varphi_1$
 field comes to somewhat exceed (by a factor of the order of 10) that of the 
 $\varphi_0$ and also somewhat exceeds that of the resonating and much 
 grown $\chi$ particle field. Even while this latter feature could be to an extent
 parameter dependent we consider a first order perturbative treatment of
 $\varphi_1$ to be inappropriate. 

  Some features are illustrated by the behaviour of $\zeta$ which
  is constant at first but then changes as the effect of the fields 
  $\varphi_1$ and/or $\chi$ becomes no longer negligible in the other
  equations of motion. Our algebraic analysis showed that $\zeta$ 
  must change when we include quadratic terms in either $\varphi_1$
  or $\chi$. Thus the necessary particle field (inhomogeneous field)
  concommitant of the classical inflaton field does by itself produce 
 changes, independent of $\chi$.

  The question is not whether these sort of effects occur but rather 
  how strong are they? 

The recent work of refs. \cite{WMLL,LLMW} finds that they are very small
by expressing the time variation of $\zeta$ in terms of the 
non-adiabatic pressure variation. This comes using the reasonable 
physical assumption - based on the approximate validity of the 
Robertson-Walker metric - that after smoothing on a cosmological 
scale below the ones of interest the spatially inhomogeneous 
variables such as $\delta\rho_k,\delta p_k$ can be treated as 
perturbations. There is no evident agreement between that 
expression for $\dot\zeta$ and Eq.(\ref{eq25}) above which has no 
assumptions on $\delta\rho_k,\delta p_k$ being perturbative.
Further differences arise because of the use, in refs.\cite{WMLL,LLMW}, 
of power spectra of $\chi^2$ and $\zeta$ in magnitude criteria;
this gives rise to expressions such as Eq.(18) of \cite{LLMW}, 
different from any in this paper, which are typified for example
by Eq.(\ref{eq22}) for $\delta p_k$.

In this paper we have taken a high momentum ('ultra-violet')
  cut-off to make the loop integrals finite. In the cosmological 
 context such ultra-violet cut-offs occur naturally as the non-zero 
 lattice spacing in numerical configuration-space calculations
 \cite{KT,KKLT}. They are also used in other momentum-space numerical 
 calculations such as those of ref.\cite{LLMW} where, as in this paper,
 the cut-off used is $H^{-1}$. The value of $H$ is roughly that of the
 preheating period. This value could be regarded as natural in this 
 context even though a cut-off is the crudest, though still-used, 
 way of regularizing in field theory. A marked change in its value
 (which acts in comparable degree both on the $\varphi_1$ 
 integrals and on the $\chi$ integrals) 
 leads to different energy densities and effective masses, and to 
 marked variations of the changes in $\zeta$ and other quantities.  

   Except for its consistent use of stochastic variables the theory
 and its parameters that we have used is neither artfully designed nor
 unusual. We consider that it points up the potential importance
 of the inflaton particle field in reheat and points to a degree of 
 caution necessary in the choice of theory and parameters if 
 that field is to be treated by first order perturbation.  

 We thank Andrew Liddle, David Lyth and Luis Mendes for some 
 conversations and Bruce Bassett for a question.

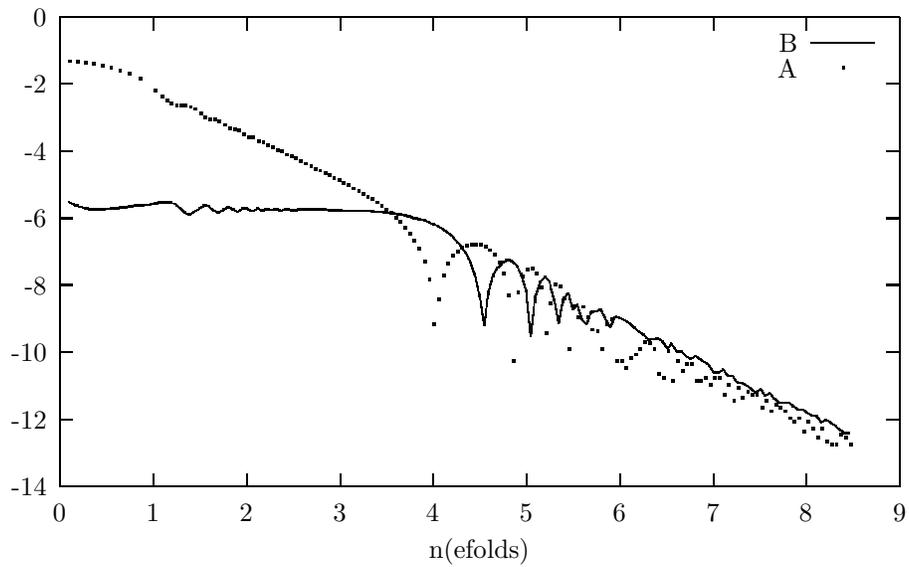
\begin{figure}[hbt]
\begin{center}
\input{FIGPLBN1.tex}
\end{center}
\caption{Logarithmic plot of energy densities of $\varphi_0$ (A)and
 $\varphi_1$ (B) versus the number of e-folds of expansion 
 in the reheat era. In all figures densities are in units
 $m^2\times m_{Planck}^2$, where $m=10^{-7}m_{Planck}$.}   
\end{figure}

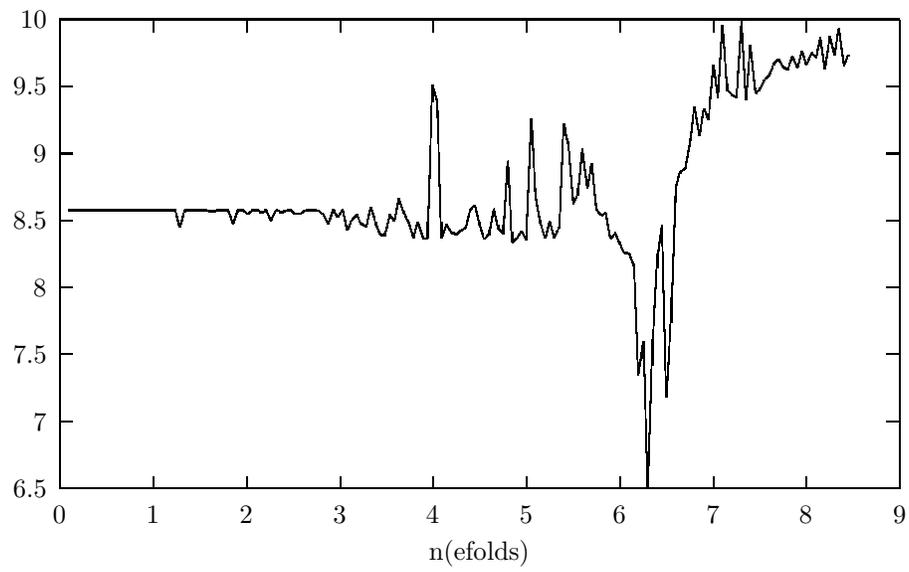
\begin{figure}[hbt]
\begin{center}
\input{FIGPLBN2.tex}
\end{center}
\caption{Plot of $\lg(\zeta)$ verss the number of e-folds of
 expansion in the reheat era.}   
\end{figure}

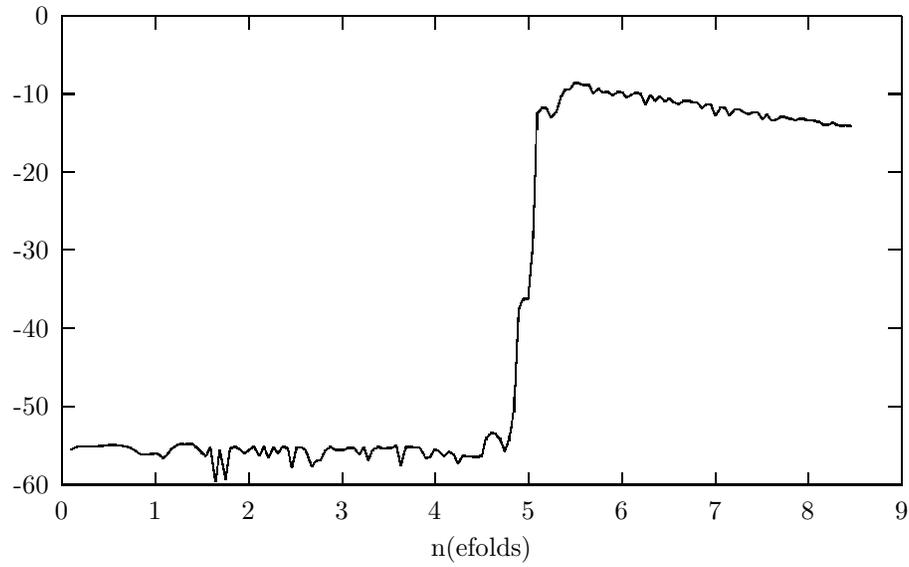
\begin{figure}[hbt]
\begin{center}
\input{FIGPLBN3.tex}
\end{center}
\caption{Logarithmic plot of the energy density of $\chi$ 
versus the number of e-folds of expansion in the reheat era.}   
\end{figure}

\begin{figure}[hbt]
\begin{center}
\input{FIGPLBN4.tex}
\end{center}
\caption{Plot of $\lg(\zeta)$ when the $\chi$ field is switched 
 off.}   
\end{figure}
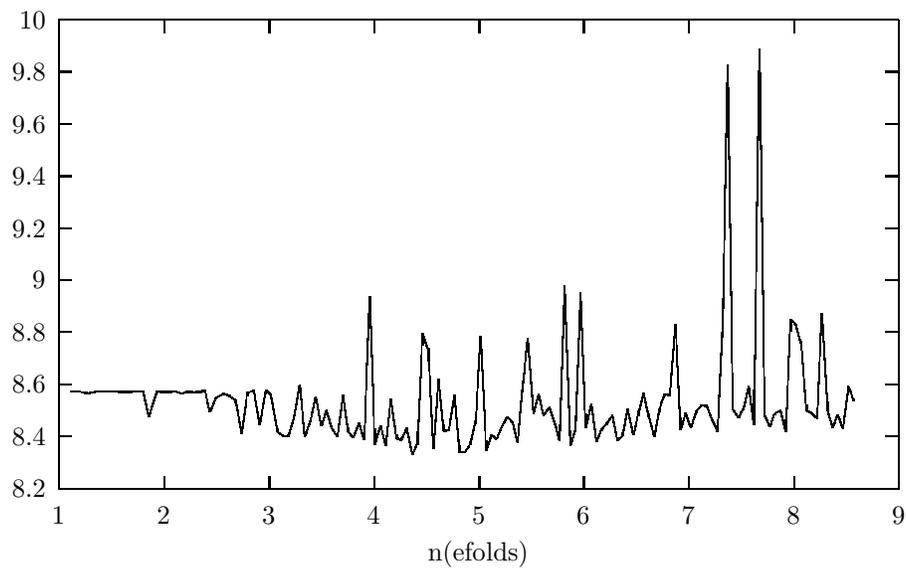

\end{document}

%% file: FIGPLBN1.tex
\setlength{\unitlength}{0.240900pt}
\ifx\plotpoint\undefined\newsavebox{\plotpoint}\fi
\sbox{\plotpoint}{\rule[-0.200pt]{0.400pt}{0.400pt}}%
\begin{picture}(1500,900)(0,0)
\font\gnuplot=cmr10 at 10pt
\gnuplot
\sbox{\plotpoint}{\rule[-0.200pt]{0.400pt}{0.400pt}}%
\put(120.0,123.0){\rule[-0.200pt]{4.818pt}{0.400pt}}
\put(100,123){\makebox(0,0)[r]{-14}}
\put(1419.0,123.0){\rule[-0.200pt]{4.818pt}{0.400pt}}
\put(120.0,228.0){\rule[-0.200pt]{4.818pt}{0.400pt}}
\put(100,228){\makebox(0,0)[r]{-12}}
\put(1419.0,228.0){\rule[-0.200pt]{4.818pt}{0.400pt}}
\put(120.0,334.0){\rule[-0.200pt]{4.818pt}{0.400pt}}
\put(100,334){\makebox(0,0)[r]{-10}}
\put(1419.0,334.0){\rule[-0.200pt]{4.818pt}{0.400pt}}
\put(120.0,439.0){\rule[-0.200pt]{4.818pt}{0.400pt}}
\put(100,439){\makebox(0,0)[r]{-8}}
\put(1419.0,439.0){\rule[-0.200pt]{4.818pt}{0.400pt}}
\put(120.0,544.0){\rule[-0.200pt]{4.818pt}{0.400pt}}
\put(100,544){\makebox(0,0)[r]{-6}}
\put(1419.0,544.0){\rule[-0.200pt]{4.818pt}{0.400pt}}
\put(120.0,649.0){\rule[-0.200pt]{4.818pt}{0.400pt}}
\put(100,649){\makebox(0,0)[r]{-4}}
\put(1419.0,649.0){\rule[-0.200pt]{4.818pt}{0.400pt}}
\put(120.0,755.0){\rule[-0.200pt]{4.818pt}{0.400pt}}
\put(100,755){\makebox(0,0)[r]{-2}}
\put(1419.0,755.0){\rule[-0.200pt]{4.818pt}{0.400pt}}
\put(120.0,860.0){\rule[-0.200pt]{4.818pt}{0.400pt}}
\put(100,860){\makebox(0,0)[r]{0}}
\put(1419.0,860.0){\rule[-0.200pt]{4.818pt}{0.400pt}}
\put(120.0,123.0){\rule[-0.200pt]{0.400pt}{4.818pt}}
\put(120,82){\makebox(0,0){0}}
\put(120.0,840.0){\rule[-0.200pt]{0.400pt}{4.818pt}}
\put(267.0,123.0){\rule[-0.200pt]{0.400pt}{4.818pt}}
\put(267,82){\makebox(0,0){1}}
\put(267.0,840.0){\rule[-0.200pt]{0.400pt}{4.818pt}}
\put(413.0,123.0){\rule[-0.200pt]{0.400pt}{4.818pt}}
\put(413,82){\makebox(0,0){2}}
\put(413.0,840.0){\rule[-0.200pt]{0.400pt}{4.818pt}}
\put(560.0,123.0){\rule[-0.200pt]{0.400pt}{4.818pt}}
\put(560,82){\makebox(0,0){3}}
\put(560.0,840.0){\rule[-0.200pt]{0.400pt}{4.818pt}}
\put(706.0,123.0){\rule[-0.200pt]{0.400pt}{4.818pt}}
\put(706,82){\makebox(0,0){4}}
\put(706.0,840.0){\rule[-0.200pt]{0.400pt}{4.818pt}}
\put(853.0,123.0){\rule[-0.200pt]{0.400pt}{4.818pt}}
\put(853,82){\makebox(0,0){5}}
\put(853.0,840.0){\rule[-0.200pt]{0.400pt}{4.818pt}}
\put(999.0,123.0){\rule[-0.200pt]{0.400pt}{4.818pt}}
\put(999,82){\makebox(0,0){6}}
\put(999.0,840.0){\rule[-0.200pt]{0.400pt}{4.818pt}}
\put(1146.0,123.0){\rule[-0.200pt]{0.400pt}{4.818pt}}
\put(1146,82){\makebox(0,0){7}}
\put(1146.0,840.0){\rule[-0.200pt]{0.400pt}{4.818pt}}
\put(1292.0,123.0){\rule[-0.200pt]{0.400pt}{4.818pt}}
\put(1292,82){\makebox(0,0){8}}
\put(1292.0,840.0){\rule[-0.200pt]{0.400pt}{4.818pt}}
\put(1439.0,123.0){\rule[-0.200pt]{0.400pt}{4.818pt}}
\put(1439,82){\makebox(0,0){9}}
\put(1439.0,840.0){\rule[-0.200pt]{0.400pt}{4.818pt}}
\put(120.0,123.0){\rule[-0.200pt]{317.747pt}{0.400pt}}
\put(1439.0,123.0){\rule[-0.200pt]{0.400pt}{177.543pt}}
\put(120.0,860.0){\rule[-0.200pt]{317.747pt}{0.400pt}}
\put(779,21){\makebox(0,0){n(efolds)}}
\put(120.0,123.0){\rule[-0.200pt]{0.400pt}{177.543pt}}
\put(1279,820){\makebox(0,0)[r]{B}}
\put(1299.0,820.0){\rule[-0.200pt]{24.090pt}{0.400pt}}
\put(134,570){\usebox{\plotpoint}}
\multiput(134.00,568.93)(1.033,-0.482){9}{\rule{0.900pt}{0.116pt}}
\multiput(134.00,569.17)(10.132,-6.000){2}{\rule{0.450pt}{0.400pt}}
\multiput(146.00,562.94)(1.797,-0.468){5}{\rule{1.400pt}{0.113pt}}
\multiput(146.00,563.17)(10.094,-4.000){2}{\rule{0.700pt}{0.400pt}}
\put(159,558.17){\rule{2.700pt}{0.400pt}}
\multiput(159.00,559.17)(7.396,-2.000){2}{\rule{1.350pt}{0.400pt}}
\put(185,557.67){\rule{3.373pt}{0.400pt}}
\multiput(185.00,557.17)(7.000,1.000){2}{\rule{1.686pt}{0.400pt}}
\put(199,558.67){\rule{3.373pt}{0.400pt}}
\multiput(199.00,558.17)(7.000,1.000){2}{\rule{1.686pt}{0.400pt}}
\put(213,560.17){\rule{3.100pt}{0.400pt}}
\multiput(213.00,559.17)(8.566,2.000){2}{\rule{1.550pt}{0.400pt}}
\put(228,562.17){\rule{3.500pt}{0.400pt}}
\multiput(228.00,561.17)(9.736,2.000){2}{\rule{1.750pt}{0.400pt}}
\multiput(245.00,564.61)(5.151,0.447){3}{\rule{3.300pt}{0.108pt}}
\multiput(245.00,563.17)(17.151,3.000){2}{\rule{1.650pt}{0.400pt}}
\put(269,567.17){\rule{2.300pt}{0.400pt}}
\multiput(269.00,566.17)(6.226,2.000){2}{\rule{1.150pt}{0.400pt}}
\put(172.0,558.0){\rule[-0.200pt]{3.132pt}{0.400pt}}
\multiput(294.00,567.95)(1.579,-0.447){3}{\rule{1.167pt}{0.108pt}}
\multiput(294.00,568.17)(5.579,-3.000){2}{\rule{0.583pt}{0.400pt}}
\multiput(302.00,564.93)(0.581,-0.482){9}{\rule{0.567pt}{0.116pt}}
\multiput(302.00,565.17)(5.824,-6.000){2}{\rule{0.283pt}{0.400pt}}
\multiput(309.00,558.93)(0.492,-0.485){11}{\rule{0.500pt}{0.117pt}}
\multiput(309.00,559.17)(5.962,-7.000){2}{\rule{0.250pt}{0.400pt}}
\multiput(316.00,551.95)(1.579,-0.447){3}{\rule{1.167pt}{0.108pt}}
\multiput(316.00,552.17)(5.579,-3.000){2}{\rule{0.583pt}{0.400pt}}
\multiput(324.00,550.60)(0.920,0.468){5}{\rule{0.800pt}{0.113pt}}
\multiput(324.00,549.17)(5.340,4.000){2}{\rule{0.400pt}{0.400pt}}
\multiput(331.00,554.59)(0.762,0.482){9}{\rule{0.700pt}{0.116pt}}
\multiput(331.00,553.17)(7.547,6.000){2}{\rule{0.350pt}{0.400pt}}
\multiput(340.00,560.60)(0.920,0.468){5}{\rule{0.800pt}{0.113pt}}
\multiput(340.00,559.17)(5.340,4.000){2}{\rule{0.400pt}{0.400pt}}
\put(347,562.67){\rule{1.686pt}{0.400pt}}
\multiput(347.00,563.17)(3.500,-1.000){2}{\rule{0.843pt}{0.400pt}}
\multiput(354.00,561.93)(0.569,-0.485){11}{\rule{0.557pt}{0.117pt}}
\multiput(354.00,562.17)(6.844,-7.000){2}{\rule{0.279pt}{0.400pt}}
\multiput(362.00,554.95)(1.355,-0.447){3}{\rule{1.033pt}{0.108pt}}
\multiput(362.00,555.17)(4.855,-3.000){2}{\rule{0.517pt}{0.400pt}}
\multiput(369.00,553.59)(0.933,0.477){7}{\rule{0.820pt}{0.115pt}}
\multiput(369.00,552.17)(7.298,5.000){2}{\rule{0.410pt}{0.400pt}}
\multiput(378.00,558.60)(0.920,0.468){5}{\rule{0.800pt}{0.113pt}}
\multiput(378.00,557.17)(5.340,4.000){2}{\rule{0.400pt}{0.400pt}}
\multiput(385.00,560.94)(1.066,-0.468){5}{\rule{0.900pt}{0.113pt}}
\multiput(385.00,561.17)(6.132,-4.000){2}{\rule{0.450pt}{0.400pt}}
\multiput(393.00,556.94)(0.920,-0.468){5}{\rule{0.800pt}{0.113pt}}
\multiput(393.00,557.17)(5.340,-4.000){2}{\rule{0.400pt}{0.400pt}}
\multiput(400.00,554.59)(0.581,0.482){9}{\rule{0.567pt}{0.116pt}}
\multiput(400.00,553.17)(5.824,6.000){2}{\rule{0.283pt}{0.400pt}}
\put(407,558.67){\rule{1.927pt}{0.400pt}}
\multiput(407.00,559.17)(4.000,-1.000){2}{\rule{0.964pt}{0.400pt}}
\multiput(415.00,557.94)(0.920,-0.468){5}{\rule{0.800pt}{0.113pt}}
\multiput(415.00,558.17)(5.340,-4.000){2}{\rule{0.400pt}{0.400pt}}
\multiput(422.00,555.59)(0.933,0.477){7}{\rule{0.820pt}{0.115pt}}
\multiput(422.00,554.17)(7.298,5.000){2}{\rule{0.410pt}{0.400pt}}
\multiput(431.00,558.94)(0.920,-0.468){5}{\rule{0.800pt}{0.113pt}}
\multiput(431.00,559.17)(5.340,-4.000){2}{\rule{0.400pt}{0.400pt}}
\put(438,556.17){\rule{1.500pt}{0.400pt}}
\multiput(438.00,555.17)(3.887,2.000){2}{\rule{0.750pt}{0.400pt}}
\put(445,556.67){\rule{1.927pt}{0.400pt}}
\multiput(445.00,557.17)(4.000,-1.000){2}{\rule{0.964pt}{0.400pt}}
\put(453,556.67){\rule{1.686pt}{0.400pt}}
\multiput(453.00,556.17)(3.500,1.000){2}{\rule{0.843pt}{0.400pt}}
\put(460,556.17){\rule{1.500pt}{0.400pt}}
\multiput(460.00,557.17)(3.887,-2.000){2}{\rule{0.750pt}{0.400pt}}
\multiput(467.00,556.61)(1.579,0.447){3}{\rule{1.167pt}{0.108pt}}
\multiput(467.00,555.17)(5.579,3.000){2}{\rule{0.583pt}{0.400pt}}
\put(475,557.17){\rule{1.500pt}{0.400pt}}
\multiput(475.00,558.17)(3.887,-2.000){2}{\rule{0.750pt}{0.400pt}}
\put(482,555.67){\rule{1.686pt}{0.400pt}}
\multiput(482.00,556.17)(3.500,-1.000){2}{\rule{0.843pt}{0.400pt}}
\put(489,556.17){\rule{1.700pt}{0.400pt}}
\multiput(489.00,555.17)(4.472,2.000){2}{\rule{0.850pt}{0.400pt}}
\put(280.0,569.0){\rule[-0.200pt]{3.373pt}{0.400pt}}
\put(535,556.67){\rule{1.686pt}{0.400pt}}
\multiput(535.00,557.17)(3.500,-1.000){2}{\rule{0.843pt}{0.400pt}}
\put(542,555.67){\rule{1.686pt}{0.400pt}}
\multiput(542.00,556.17)(3.500,-1.000){2}{\rule{0.843pt}{0.400pt}}
\put(549,555.67){\rule{1.927pt}{0.400pt}}
\multiput(549.00,555.17)(4.000,1.000){2}{\rule{0.964pt}{0.400pt}}
\put(557,555.67){\rule{1.686pt}{0.400pt}}
\multiput(557.00,556.17)(3.500,-1.000){2}{\rule{0.843pt}{0.400pt}}
\put(497.0,558.0){\rule[-0.200pt]{9.154pt}{0.400pt}}
\put(595,554.67){\rule{1.686pt}{0.400pt}}
\multiput(595.00,555.17)(3.500,-1.000){2}{\rule{0.843pt}{0.400pt}}
\put(564.0,556.0){\rule[-0.200pt]{7.468pt}{0.400pt}}
\put(609,553.67){\rule{1.927pt}{0.400pt}}
\multiput(609.00,554.17)(4.000,-1.000){2}{\rule{0.964pt}{0.400pt}}
\put(602.0,555.0){\rule[-0.200pt]{1.686pt}{0.400pt}}
\put(624,552.67){\rule{1.686pt}{0.400pt}}
\multiput(624.00,553.17)(3.500,-1.000){2}{\rule{0.843pt}{0.400pt}}
\put(631,551.67){\rule{1.927pt}{0.400pt}}
\multiput(631.00,552.17)(4.000,-1.000){2}{\rule{0.964pt}{0.400pt}}
\put(639,550.67){\rule{1.686pt}{0.400pt}}
\multiput(639.00,551.17)(3.500,-1.000){2}{\rule{0.843pt}{0.400pt}}
\put(646,549.67){\rule{1.686pt}{0.400pt}}
\multiput(646.00,550.17)(3.500,-1.000){2}{\rule{0.843pt}{0.400pt}}
\put(653,548.67){\rule{1.927pt}{0.400pt}}
\multiput(653.00,549.17)(4.000,-1.000){2}{\rule{0.964pt}{0.400pt}}
\put(661,547.17){\rule{1.500pt}{0.400pt}}
\multiput(661.00,548.17)(3.887,-2.000){2}{\rule{0.750pt}{0.400pt}}
\put(668,545.17){\rule{1.500pt}{0.400pt}}
\multiput(668.00,546.17)(3.887,-2.000){2}{\rule{0.750pt}{0.400pt}}
\put(675,543.17){\rule{1.700pt}{0.400pt}}
\multiput(675.00,544.17)(4.472,-2.000){2}{\rule{0.850pt}{0.400pt}}
\put(683,541.17){\rule{1.900pt}{0.400pt}}
\multiput(683.00,542.17)(5.056,-2.000){2}{\rule{0.950pt}{0.400pt}}
\multiput(692.00,539.95)(1.355,-0.447){3}{\rule{1.033pt}{0.108pt}}
\multiput(692.00,540.17)(4.855,-3.000){2}{\rule{0.517pt}{0.400pt}}
\multiput(699.00,536.95)(1.355,-0.447){3}{\rule{1.033pt}{0.108pt}}
\multiput(699.00,537.17)(4.855,-3.000){2}{\rule{0.517pt}{0.400pt}}
\multiput(706.00,533.95)(1.579,-0.447){3}{\rule{1.167pt}{0.108pt}}
\multiput(706.00,534.17)(5.579,-3.000){2}{\rule{0.583pt}{0.400pt}}
\multiput(714.00,530.93)(0.710,-0.477){7}{\rule{0.660pt}{0.115pt}}
\multiput(714.00,531.17)(5.630,-5.000){2}{\rule{0.330pt}{0.400pt}}
\multiput(721.00,525.94)(0.920,-0.468){5}{\rule{0.800pt}{0.113pt}}
\multiput(721.00,526.17)(5.340,-4.000){2}{\rule{0.400pt}{0.400pt}}
\multiput(728.00,521.93)(0.671,-0.482){9}{\rule{0.633pt}{0.116pt}}
\multiput(728.00,522.17)(6.685,-6.000){2}{\rule{0.317pt}{0.400pt}}
\multiput(736.59,514.69)(0.485,-0.569){11}{\rule{0.117pt}{0.557pt}}
\multiput(735.17,515.84)(7.000,-6.844){2}{\rule{0.400pt}{0.279pt}}
\multiput(743.59,506.69)(0.485,-0.569){11}{\rule{0.117pt}{0.557pt}}
\multiput(742.17,507.84)(7.000,-6.844){2}{\rule{0.400pt}{0.279pt}}
\multiput(750.59,498.30)(0.488,-0.692){13}{\rule{0.117pt}{0.650pt}}
\multiput(749.17,499.65)(8.000,-9.651){2}{\rule{0.400pt}{0.325pt}}
\multiput(758.59,486.26)(0.485,-1.026){11}{\rule{0.117pt}{0.900pt}}
\multiput(757.17,488.13)(7.000,-12.132){2}{\rule{0.400pt}{0.450pt}}
\multiput(765.59,470.84)(0.485,-1.484){11}{\rule{0.117pt}{1.243pt}}
\multiput(764.17,473.42)(7.000,-17.420){2}{\rule{0.400pt}{0.621pt}}
\multiput(772.59,448.32)(0.488,-2.277){13}{\rule{0.117pt}{1.850pt}}
\multiput(771.17,452.16)(8.000,-31.160){2}{\rule{0.400pt}{0.925pt}}
\multiput(780.59,409.91)(0.485,-3.391){11}{\rule{0.117pt}{2.671pt}}
\multiput(779.17,415.46)(7.000,-39.455){2}{\rule{0.400pt}{1.336pt}}
\multiput(787.59,376.00)(0.485,4.078){11}{\rule{0.117pt}{3.186pt}}
\multiput(786.17,376.00)(7.000,47.388){2}{\rule{0.400pt}{1.593pt}}
\multiput(794.59,430.00)(0.485,1.865){11}{\rule{0.117pt}{1.529pt}}
\multiput(793.17,430.00)(7.000,21.827){2}{\rule{0.400pt}{0.764pt}}
\multiput(801.59,455.00)(0.488,0.824){13}{\rule{0.117pt}{0.750pt}}
\multiput(800.17,455.00)(8.000,11.443){2}{\rule{0.400pt}{0.375pt}}
\multiput(809.00,468.59)(0.492,0.485){11}{\rule{0.500pt}{0.117pt}}
\multiput(809.00,467.17)(5.962,7.000){2}{\rule{0.250pt}{0.400pt}}
\multiput(816.00,475.61)(1.355,0.447){3}{\rule{1.033pt}{0.108pt}}
\multiput(816.00,474.17)(4.855,3.000){2}{\rule{0.517pt}{0.400pt}}
\put(823,476.17){\rule{1.700pt}{0.400pt}}
\multiput(823.00,477.17)(4.472,-2.000){2}{\rule{0.850pt}{0.400pt}}
\multiput(831.00,474.93)(0.581,-0.482){9}{\rule{0.567pt}{0.116pt}}
\multiput(831.00,475.17)(5.824,-6.000){2}{\rule{0.283pt}{0.400pt}}
\multiput(838.59,466.26)(0.485,-1.026){11}{\rule{0.117pt}{0.900pt}}
\multiput(837.17,468.13)(7.000,-12.132){2}{\rule{0.400pt}{0.450pt}}
\multiput(845.59,449.98)(0.488,-1.748){13}{\rule{0.117pt}{1.450pt}}
\multiput(844.17,452.99)(8.000,-23.990){2}{\rule{0.400pt}{0.725pt}}
\multiput(853.59,411.98)(0.485,-5.298){11}{\rule{0.117pt}{4.100pt}}
\multiput(852.17,420.49)(7.000,-61.490){2}{\rule{0.400pt}{2.050pt}}
\multiput(860.59,359.00)(0.485,4.535){11}{\rule{0.117pt}{3.529pt}}
\multiput(859.17,359.00)(7.000,52.676){2}{\rule{0.400pt}{1.764pt}}
\multiput(867.59,419.00)(0.488,1.682){13}{\rule{0.117pt}{1.400pt}}
\multiput(866.17,419.00)(8.000,23.094){2}{\rule{0.400pt}{0.700pt}}
\multiput(875.00,445.59)(0.492,0.485){11}{\rule{0.500pt}{0.117pt}}
\multiput(875.00,444.17)(5.962,7.000){2}{\rule{0.250pt}{0.400pt}}
\multiput(882.00,450.93)(0.492,-0.485){11}{\rule{0.500pt}{0.117pt}}
\multiput(882.00,451.17)(5.962,-7.000){2}{\rule{0.250pt}{0.400pt}}
\multiput(889.59,438.36)(0.488,-1.947){13}{\rule{0.117pt}{1.600pt}}
\multiput(888.17,441.68)(8.000,-26.679){2}{\rule{0.400pt}{0.800pt}}
\multiput(897.59,405.81)(0.485,-2.781){11}{\rule{0.117pt}{2.214pt}}
\multiput(896.17,410.40)(7.000,-32.404){2}{\rule{0.400pt}{1.107pt}}
\multiput(904.59,378.00)(0.485,2.933){11}{\rule{0.117pt}{2.329pt}}
\multiput(903.17,378.00)(7.000,34.167){2}{\rule{0.400pt}{1.164pt}}
\multiput(911.59,417.00)(0.488,0.626){13}{\rule{0.117pt}{0.600pt}}
\multiput(910.17,417.00)(8.000,8.755){2}{\rule{0.400pt}{0.300pt}}
\multiput(919.59,420.42)(0.485,-1.942){11}{\rule{0.117pt}{1.586pt}}
\multiput(918.17,423.71)(7.000,-22.709){2}{\rule{0.400pt}{0.793pt}}
\multiput(926.00,401.59)(0.492,0.485){11}{\rule{0.500pt}{0.117pt}}
\multiput(926.00,400.17)(5.962,7.000){2}{\rule{0.250pt}{0.400pt}}
\multiput(933.59,402.60)(0.488,-1.550){13}{\rule{0.117pt}{1.300pt}}
\multiput(932.17,405.30)(8.000,-21.302){2}{\rule{0.400pt}{0.650pt}}
\multiput(941.00,382.93)(0.581,-0.482){9}{\rule{0.567pt}{0.116pt}}
\multiput(941.00,383.17)(5.824,-6.000){2}{\rule{0.283pt}{0.400pt}}
\multiput(948.59,378.00)(0.485,1.408){11}{\rule{0.117pt}{1.186pt}}
\multiput(947.17,378.00)(7.000,16.539){2}{\rule{0.400pt}{0.593pt}}
\put(955,396.67){\rule{1.927pt}{0.400pt}}
\multiput(955.00,396.17)(4.000,1.000){2}{\rule{0.964pt}{0.400pt}}
\multiput(963.00,398.61)(1.355,0.447){3}{\rule{1.033pt}{0.108pt}}
\multiput(963.00,397.17)(4.855,3.000){2}{\rule{0.517pt}{0.400pt}}
\multiput(970.59,397.26)(0.485,-1.026){11}{\rule{0.117pt}{0.900pt}}
\multiput(969.17,399.13)(7.000,-12.132){2}{\rule{0.400pt}{0.450pt}}
\multiput(977.59,383.68)(0.488,-0.890){13}{\rule{0.117pt}{0.800pt}}
\multiput(976.17,385.34)(8.000,-12.340){2}{\rule{0.400pt}{0.400pt}}
\multiput(985.59,373.00)(0.488,1.088){13}{\rule{0.117pt}{0.950pt}}
\multiput(984.17,373.00)(8.000,15.028){2}{\rule{0.400pt}{0.475pt}}
\multiput(993.00,388.95)(1.579,-0.447){3}{\rule{1.167pt}{0.108pt}}
\multiput(993.00,389.17)(5.579,-3.000){2}{\rule{0.583pt}{0.400pt}}
\multiput(1001.00,385.95)(1.355,-0.447){3}{\rule{1.033pt}{0.108pt}}
\multiput(1001.00,386.17)(4.855,-3.000){2}{\rule{0.517pt}{0.400pt}}
\multiput(1008.00,382.93)(0.710,-0.477){7}{\rule{0.660pt}{0.115pt}}
\multiput(1008.00,383.17)(5.630,-5.000){2}{\rule{0.330pt}{0.400pt}}
\multiput(1015.00,377.93)(0.569,-0.485){11}{\rule{0.557pt}{0.117pt}}
\multiput(1015.00,378.17)(6.844,-7.000){2}{\rule{0.279pt}{0.400pt}}
\multiput(1023.00,370.93)(0.581,-0.482){9}{\rule{0.567pt}{0.116pt}}
\multiput(1023.00,371.17)(5.824,-6.000){2}{\rule{0.283pt}{0.400pt}}
\multiput(1030.00,364.93)(0.710,-0.477){7}{\rule{0.660pt}{0.115pt}}
\multiput(1030.00,365.17)(5.630,-5.000){2}{\rule{0.330pt}{0.400pt}}
\multiput(1037.00,359.93)(0.494,-0.488){13}{\rule{0.500pt}{0.117pt}}
\multiput(1037.00,360.17)(6.962,-8.000){2}{\rule{0.250pt}{0.400pt}}
\put(1045,352.67){\rule{1.686pt}{0.400pt}}
\multiput(1045.00,352.17)(3.500,1.000){2}{\rule{0.843pt}{0.400pt}}
\put(1052,353.67){\rule{1.686pt}{0.400pt}}
\multiput(1052.00,353.17)(3.500,1.000){2}{\rule{0.843pt}{0.400pt}}
\multiput(1059.00,353.94)(1.066,-0.468){5}{\rule{0.900pt}{0.113pt}}
\multiput(1059.00,354.17)(6.132,-4.000){2}{\rule{0.450pt}{0.400pt}}
\multiput(1067.59,347.98)(0.485,-0.798){11}{\rule{0.117pt}{0.729pt}}
\multiput(1066.17,349.49)(7.000,-9.488){2}{\rule{0.400pt}{0.364pt}}
\multiput(1074.00,340.59)(0.492,0.485){11}{\rule{0.500pt}{0.117pt}}
\multiput(1074.00,339.17)(5.962,7.000){2}{\rule{0.250pt}{0.400pt}}
\multiput(1081.59,344.09)(0.488,-0.758){13}{\rule{0.117pt}{0.700pt}}
\multiput(1080.17,345.55)(8.000,-10.547){2}{\rule{0.400pt}{0.350pt}}
\put(617.0,554.0){\rule[-0.200pt]{1.686pt}{0.400pt}}
\multiput(1096.00,333.93)(0.492,-0.485){11}{\rule{0.500pt}{0.117pt}}
\multiput(1096.00,334.17)(5.962,-7.000){2}{\rule{0.250pt}{0.400pt}}
\multiput(1103.00,326.93)(0.821,-0.477){7}{\rule{0.740pt}{0.115pt}}
\multiput(1103.00,327.17)(6.464,-5.000){2}{\rule{0.370pt}{0.400pt}}
\multiput(1111.00,323.59)(0.710,0.477){7}{\rule{0.660pt}{0.115pt}}
\multiput(1111.00,322.17)(5.630,5.000){2}{\rule{0.330pt}{0.400pt}}
\multiput(1118.00,326.93)(0.710,-0.477){7}{\rule{0.660pt}{0.115pt}}
\multiput(1118.00,327.17)(5.630,-5.000){2}{\rule{0.330pt}{0.400pt}}
\multiput(1125.00,321.93)(0.821,-0.477){7}{\rule{0.740pt}{0.115pt}}
\multiput(1125.00,322.17)(6.464,-5.000){2}{\rule{0.370pt}{0.400pt}}
\multiput(1133.00,316.93)(0.710,-0.477){7}{\rule{0.660pt}{0.115pt}}
\multiput(1133.00,317.17)(5.630,-5.000){2}{\rule{0.330pt}{0.400pt}}
\multiput(1140.59,309.98)(0.485,-0.798){11}{\rule{0.117pt}{0.729pt}}
\multiput(1139.17,311.49)(7.000,-9.488){2}{\rule{0.400pt}{0.364pt}}
\put(1089.0,335.0){\rule[-0.200pt]{1.686pt}{0.400pt}}
\multiput(1155.00,302.59)(0.710,0.477){7}{\rule{0.660pt}{0.115pt}}
\multiput(1155.00,301.17)(5.630,5.000){2}{\rule{0.330pt}{0.400pt}}
\multiput(1162.59,304.21)(0.485,-0.721){11}{\rule{0.117pt}{0.671pt}}
\multiput(1161.17,305.61)(7.000,-8.606){2}{\rule{0.400pt}{0.336pt}}
\put(1147.0,302.0){\rule[-0.200pt]{1.927pt}{0.400pt}}
\multiput(1177.59,293.98)(0.485,-0.798){11}{\rule{0.117pt}{0.729pt}}
\multiput(1176.17,295.49)(7.000,-9.488){2}{\rule{0.400pt}{0.364pt}}
\put(1169.0,297.0){\rule[-0.200pt]{1.927pt}{0.400pt}}
\multiput(1191.00,284.93)(0.821,-0.477){7}{\rule{0.740pt}{0.115pt}}
\multiput(1191.00,285.17)(6.464,-5.000){2}{\rule{0.370pt}{0.400pt}}
\multiput(1199.00,279.93)(0.710,-0.477){7}{\rule{0.660pt}{0.115pt}}
\multiput(1199.00,280.17)(5.630,-5.000){2}{\rule{0.330pt}{0.400pt}}
\multiput(1206.00,274.93)(0.581,-0.482){9}{\rule{0.567pt}{0.116pt}}
\multiput(1206.00,275.17)(5.824,-6.000){2}{\rule{0.283pt}{0.400pt}}
\multiput(1213.00,270.59)(0.671,0.482){9}{\rule{0.633pt}{0.116pt}}
\multiput(1213.00,269.17)(6.685,6.000){2}{\rule{0.317pt}{0.400pt}}
\multiput(1221.59,272.98)(0.485,-0.798){11}{\rule{0.117pt}{0.729pt}}
\multiput(1220.17,274.49)(7.000,-9.488){2}{\rule{0.400pt}{0.364pt}}
\multiput(1228.00,265.59)(0.710,0.477){7}{\rule{0.660pt}{0.115pt}}
\multiput(1228.00,264.17)(5.630,5.000){2}{\rule{0.330pt}{0.400pt}}
\multiput(1235.59,267.51)(0.488,-0.626){13}{\rule{0.117pt}{0.600pt}}
\multiput(1234.17,268.75)(8.000,-8.755){2}{\rule{0.400pt}{0.300pt}}
\multiput(1243.00,258.93)(0.710,-0.477){7}{\rule{0.660pt}{0.115pt}}
\multiput(1243.00,259.17)(5.630,-5.000){2}{\rule{0.330pt}{0.400pt}}
\put(1184.0,286.0){\rule[-0.200pt]{1.686pt}{0.400pt}}
\multiput(1265.00,253.93)(0.581,-0.482){9}{\rule{0.567pt}{0.116pt}}
\multiput(1265.00,254.17)(5.824,-6.000){2}{\rule{0.283pt}{0.400pt}}
\multiput(1272.00,247.93)(0.710,-0.477){7}{\rule{0.660pt}{0.115pt}}
\multiput(1272.00,248.17)(5.630,-5.000){2}{\rule{0.330pt}{0.400pt}}
\put(1250.0,255.0){\rule[-0.200pt]{3.613pt}{0.400pt}}
\multiput(1287.00,242.93)(0.710,-0.477){7}{\rule{0.660pt}{0.115pt}}
\multiput(1287.00,243.17)(5.630,-5.000){2}{\rule{0.330pt}{0.400pt}}
\multiput(1294.00,237.93)(0.710,-0.477){7}{\rule{0.660pt}{0.115pt}}
\multiput(1294.00,238.17)(5.630,-5.000){2}{\rule{0.330pt}{0.400pt}}
\put(1279.0,244.0){\rule[-0.200pt]{1.927pt}{0.400pt}}
\multiput(1309.59,230.98)(0.485,-0.798){11}{\rule{0.117pt}{0.729pt}}
\multiput(1308.17,232.49)(7.000,-9.488){2}{\rule{0.400pt}{0.364pt}}
\multiput(1316.00,223.59)(0.710,0.477){7}{\rule{0.660pt}{0.115pt}}
\multiput(1316.00,222.17)(5.630,5.000){2}{\rule{0.330pt}{0.400pt}}
\multiput(1323.00,226.93)(0.821,-0.477){7}{\rule{0.740pt}{0.115pt}}
\multiput(1323.00,227.17)(6.464,-5.000){2}{\rule{0.370pt}{0.400pt}}
\multiput(1331.00,221.93)(0.710,-0.477){7}{\rule{0.660pt}{0.115pt}}
\multiput(1331.00,222.17)(5.630,-5.000){2}{\rule{0.330pt}{0.400pt}}
\multiput(1338.00,216.93)(0.581,-0.482){9}{\rule{0.567pt}{0.116pt}}
\multiput(1338.00,217.17)(5.824,-6.000){2}{\rule{0.283pt}{0.400pt}}
\multiput(1345.00,210.93)(0.821,-0.477){7}{\rule{0.740pt}{0.115pt}}
\multiput(1345.00,211.17)(6.464,-5.000){2}{\rule{0.370pt}{0.400pt}}
\put(1301.0,234.0){\rule[-0.200pt]{1.927pt}{0.400pt}}
\put(1353.0,207.0){\rule[-0.200pt]{1.686pt}{0.400pt}}
\put(1279,779){\makebox(0,0)[r]{A}}
\put(1349,779){\rule{1pt}{1pt}}
\put(134,789){\rule{1pt}{1pt}}
\put(146,788){\rule{1pt}{1pt}}
\put(159,787){\rule{1pt}{1pt}}
\put(172,785){\rule{1pt}{1pt}}
\put(185,782){\rule{1pt}{1pt}}
\put(199,779){\rule{1pt}{1pt}}
\put(213,774){\rule{1pt}{1pt}}
\put(228,769){\rule{1pt}{1pt}}
\put(245,761){\rule{1pt}{1pt}}
\put(269,743){\rule{1pt}{1pt}}
\put(280,734){\rule{1pt}{1pt}}
\put(287,727){\rule{1pt}{1pt}}
\put(294,722){\rule{1pt}{1pt}}
\put(302,720){\rule{1pt}{1pt}}
\put(309,719){\rule{1pt}{1pt}}
\put(316,719){\rule{1pt}{1pt}}
\put(324,717){\rule{1pt}{1pt}}
\put(331,714){\rule{1pt}{1pt}}
\put(340,707){\rule{1pt}{1pt}}
\put(347,700){\rule{1pt}{1pt}}
\put(354,697){\rule{1pt}{1pt}}
\put(362,697){\rule{1pt}{1pt}}
\put(369,695){\rule{1pt}{1pt}}
\put(378,689){\rule{1pt}{1pt}}
\put(385,683){\rule{1pt}{1pt}}
\put(393,682){\rule{1pt}{1pt}}
\put(400,680){\rule{1pt}{1pt}}
\put(407,674){\rule{1pt}{1pt}}
\put(415,670){\rule{1pt}{1pt}}
\put(422,669){\rule{1pt}{1pt}}
\put(431,663){\rule{1pt}{1pt}}
\put(438,661){\rule{1pt}{1pt}}
\put(445,657){\rule{1pt}{1pt}}
\put(453,654){\rule{1pt}{1pt}}
\put(460,649){\rule{1pt}{1pt}}
\put(467,647){\rule{1pt}{1pt}}
\put(475,642){\rule{1pt}{1pt}}
\put(482,640){\rule{1pt}{1pt}}
\put(489,636){\rule{1pt}{1pt}}
\put(497,632){\rule{1pt}{1pt}}
\put(504,628){\rule{1pt}{1pt}}
\put(513,624){\rule{1pt}{1pt}}
\put(520,620){\rule{1pt}{1pt}}
\put(527,617){\rule{1pt}{1pt}}
\put(535,613){\rule{1pt}{1pt}}
\put(542,610){\rule{1pt}{1pt}}
\put(549,606){\rule{1pt}{1pt}}
\put(557,602){\rule{1pt}{1pt}}
\put(564,598){\rule{1pt}{1pt}}
\put(571,594){\rule{1pt}{1pt}}
\put(580,589){\rule{1pt}{1pt}}
\put(588,585){\rule{1pt}{1pt}}
\put(595,580){\rule{1pt}{1pt}}
\put(602,576){\rule{1pt}{1pt}}
\put(609,572){\rule{1pt}{1pt}}
\put(617,567){\rule{1pt}{1pt}}
\put(624,561){\rule{1pt}{1pt}}
\put(631,556){\rule{1pt}{1pt}}
\put(639,550){\rule{1pt}{1pt}}
\put(646,543){\rule{1pt}{1pt}}
\put(653,536){\rule{1pt}{1pt}}
\put(661,527){\rule{1pt}{1pt}}
\put(668,518){\rule{1pt}{1pt}}
\put(675,507){\rule{1pt}{1pt}}
\put(683,494){\rule{1pt}{1pt}}
\put(692,474){\rule{1pt}{1pt}}
\put(699,446){\rule{1pt}{1pt}}
\put(706,376){\rule{1pt}{1pt}}
\put(714,415){\rule{1pt}{1pt}}
\put(721,452){\rule{1pt}{1pt}}
\put(728,471){\rule{1pt}{1pt}}
\put(736,483){\rule{1pt}{1pt}}
\put(743,490){\rule{1pt}{1pt}}
\put(750,496){\rule{1pt}{1pt}}
\put(758,499){\rule{1pt}{1pt}}
\put(765,500){\rule{1pt}{1pt}}
\put(772,501){\rule{1pt}{1pt}}
\put(780,500){\rule{1pt}{1pt}}
\put(787,497){\rule{1pt}{1pt}}
\put(794,492){\rule{1pt}{1pt}}
\put(801,485){\rule{1pt}{1pt}}
\put(809,473){\rule{1pt}{1pt}}
\put(816,455){\rule{1pt}{1pt}}
\put(823,421){\rule{1pt}{1pt}}
\put(831,318){\rule{1pt}{1pt}}
\put(838,426){\rule{1pt}{1pt}}
\put(845,451){\rule{1pt}{1pt}}
\put(853,461){\rule{1pt}{1pt}}
\put(860,463){\rule{1pt}{1pt}}
\put(867,455){\rule{1pt}{1pt}}
\put(875,434){\rule{1pt}{1pt}}
\put(882,362){\rule{1pt}{1pt}}
\put(889,409){\rule{1pt}{1pt}}
\put(897,438){\rule{1pt}{1pt}}
\put(904,435){\rule{1pt}{1pt}}
\put(911,416){\rule{1pt}{1pt}}
\put(919,336){\rule{1pt}{1pt}}
\put(926,404){\rule{1pt}{1pt}}
\put(933,386){\rule{1pt}{1pt}}
\put(941,403){\rule{1pt}{1pt}}
\put(948,386){\rule{1pt}{1pt}}
\put(955,367){\rule{1pt}{1pt}}
\put(963,365){\rule{1pt}{1pt}}
\put(970,337){\rule{1pt}{1pt}}
\put(977,375){\rule{1pt}{1pt}}
\put(985,383){\rule{1pt}{1pt}}
\put(993,318){\rule{1pt}{1pt}}
\put(1001,318){\rule{1pt}{1pt}}
\put(1008,307){\rule{1pt}{1pt}}
\put(1015,323){\rule{1pt}{1pt}}
\put(1023,328){\rule{1pt}{1pt}}
\put(1030,337){\rule{1pt}{1pt}}
\put(1037,347){\rule{1pt}{1pt}}
\put(1045,346){\rule{1pt}{1pt}}
\put(1052,336){\rule{1pt}{1pt}}
\put(1059,297){\rule{1pt}{1pt}}
\put(1067,291){\rule{1pt}{1pt}}
\put(1074,334){\rule{1pt}{1pt}}
\put(1081,286){\rule{1pt}{1pt}}
\put(1089,318){\rule{1pt}{1pt}}
\put(1096,302){\rule{1pt}{1pt}}
\put(1103,313){\rule{1pt}{1pt}}
\put(1111,313){\rule{1pt}{1pt}}
\put(1118,286){\rule{1pt}{1pt}}
\put(1125,286){\rule{1pt}{1pt}}
\put(1133,291){\rule{1pt}{1pt}}
\put(1140,281){\rule{1pt}{1pt}}
\put(1147,291){\rule{1pt}{1pt}}
\put(1155,291){\rule{1pt}{1pt}}
\put(1162,265){\rule{1pt}{1pt}}
\put(1169,281){\rule{1pt}{1pt}}
\put(1177,255){\rule{1pt}{1pt}}
\put(1184,276){\rule{1pt}{1pt}}
\put(1191,260){\rule{1pt}{1pt}}
\put(1199,270){\rule{1pt}{1pt}}
\put(1206,265){\rule{1pt}{1pt}}
\put(1213,265){\rule{1pt}{1pt}}
\put(1221,244){\rule{1pt}{1pt}}
\put(1228,255){\rule{1pt}{1pt}}
\put(1235,239){\rule{1pt}{1pt}}
\put(1243,249){\rule{1pt}{1pt}}
\put(1250,244){\rule{1pt}{1pt}}
\put(1257,239){\rule{1pt}{1pt}}
\put(1265,228){\rule{1pt}{1pt}}
\put(1272,223){\rule{1pt}{1pt}}
\put(1279,228){\rule{1pt}{1pt}}
\put(1287,207){\rule{1pt}{1pt}}
\put(1294,223){\rule{1pt}{1pt}}
\put(1301,212){\rule{1pt}{1pt}}
\put(1309,197){\rule{1pt}{1pt}}
\put(1316,212){\rule{1pt}{1pt}}
\put(1323,191){\rule{1pt}{1pt}}
\put(1331,186){\rule{1pt}{1pt}}
\put(1338,186){\rule{1pt}{1pt}}
\put(1345,202){\rule{1pt}{1pt}}
\put(1353,197){\rule{1pt}{1pt}}
\put(1360,186){\rule{1pt}{1pt}}
\end{picture}

%% file: FIGPLBN2.tex
\setlength{\unitlength}{0.240900pt}
\ifx\plotpoint\undefined\newsavebox{\plotpoint}\fi
\begin{picture}(1500,900)(0,0)
\font\gnuplot=cmr10 at 10pt
\gnuplot
\sbox{\plotpoint}{\rule[-0.200pt]{0.400pt}{0.400pt}}%
\put(120.0,123.0){\rule[-0.200pt]{4.818pt}{0.400pt}}
\put(100,123){\makebox(0,0)[r]{6.5}}
\put(1419.0,123.0){\rule[-0.200pt]{4.818pt}{0.400pt}}
\put(120.0,228.0){\rule[-0.200pt]{4.818pt}{0.400pt}}
\put(100,228){\makebox(0,0)[r]{7}}
\put(1419.0,228.0){\rule[-0.200pt]{4.818pt}{0.400pt}}
\put(120.0,334.0){\rule[-0.200pt]{4.818pt}{0.400pt}}
\put(100,334){\makebox(0,0)[r]{7.5}}
\put(1419.0,334.0){\rule[-0.200pt]{4.818pt}{0.400pt}}
\put(120.0,439.0){\rule[-0.200pt]{4.818pt}{0.400pt}}
\put(100,439){\makebox(0,0)[r]{8}}
\put(1419.0,439.0){\rule[-0.200pt]{4.818pt}{0.400pt}}
\put(120.0,544.0){\rule[-0.200pt]{4.818pt}{0.400pt}}
\put(100,544){\makebox(0,0)[r]{8.5}}
\put(1419.0,544.0){\rule[-0.200pt]{4.818pt}{0.400pt}}
\put(120.0,649.0){\rule[-0.200pt]{4.818pt}{0.400pt}}
\put(100,649){\makebox(0,0)[r]{9}}
\put(1419.0,649.0){\rule[-0.200pt]{4.818pt}{0.400pt}}
\put(120.0,755.0){\rule[-0.200pt]{4.818pt}{0.400pt}}
\put(100,755){\makebox(0,0)[r]{9.5}}
\put(1419.0,755.0){\rule[-0.200pt]{4.818pt}{0.400pt}}
\put(120.0,860.0){\rule[-0.200pt]{4.818pt}{0.400pt}}
\put(100,860){\makebox(0,0)[r]{10}}
\put(1419.0,860.0){\rule[-0.200pt]{4.818pt}{0.400pt}}
\put(120.0,123.0){\rule[-0.200pt]{0.400pt}{4.818pt}}
\put(120,82){\makebox(0,0){0}}
\put(120.0,840.0){\rule[-0.200pt]{0.400pt}{4.818pt}}
\put(267.0,123.0){\rule[-0.200pt]{0.400pt}{4.818pt}}
\put(267,82){\makebox(0,0){1}}
\put(267.0,840.0){\rule[-0.200pt]{0.400pt}{4.818pt}}
\put(413.0,123.0){\rule[-0.200pt]{0.400pt}{4.818pt}}
\put(413,82){\makebox(0,0){2}}
\put(413.0,840.0){\rule[-0.200pt]{0.400pt}{4.818pt}}
\put(560.0,123.0){\rule[-0.200pt]{0.400pt}{4.818pt}}
\put(560,82){\makebox(0,0){3}}
\put(560.0,840.0){\rule[-0.200pt]{0.400pt}{4.818pt}}
\put(706.0,123.0){\rule[-0.200pt]{0.400pt}{4.818pt}}
\put(706,82){\makebox(0,0){4}}
\put(706.0,840.0){\rule[-0.200pt]{0.400pt}{4.818pt}}
\put(853.0,123.0){\rule[-0.200pt]{0.400pt}{4.818pt}}
\put(853,82){\makebox(0,0){5}}
\put(853.0,840.0){\rule[-0.200pt]{0.400pt}{4.818pt}}
\put(999.0,123.0){\rule[-0.200pt]{0.400pt}{4.818pt}}
\put(999,82){\makebox(0,0){6}}
\put(999.0,840.0){\rule[-0.200pt]{0.400pt}{4.818pt}}
\put(1146.0,123.0){\rule[-0.200pt]{0.400pt}{4.818pt}}
\put(1146,82){\makebox(0,0){7}}
\put(1146.0,840.0){\rule[-0.200pt]{0.400pt}{4.818pt}}
\put(1292.0,123.0){\rule[-0.200pt]{0.400pt}{4.818pt}}
\put(1292,82){\makebox(0,0){8}}
\put(1292.0,840.0){\rule[-0.200pt]{0.400pt}{4.818pt}}
\put(1439.0,123.0){\rule[-0.200pt]{0.400pt}{4.818pt}}
\put(1439,82){\makebox(0,0){9}}
\put(1439.0,840.0){\rule[-0.200pt]{0.400pt}{4.818pt}}
\put(120.0,123.0){\rule[-0.200pt]{317.747pt}{0.400pt}}
\put(1439.0,123.0){\rule[-0.200pt]{0.400pt}{177.543pt}}
\put(120.0,860.0){\rule[-0.200pt]{317.747pt}{0.400pt}}
\put(779,21){\makebox(0,0){n(efolds)}}
\put(120.0,123.0){\rule[-0.200pt]{0.400pt}{177.543pt}}
\put(134,560){\usebox{\plotpoint}}
\multiput(302.59,553.18)(0.485,-2.018){11}{\rule{0.117pt}{1.643pt}}
\multiput(301.17,556.59)(7.000,-23.590){2}{\rule{0.400pt}{0.821pt}}
\multiput(309.59,533.00)(0.488,1.748){13}{\rule{0.117pt}{1.450pt}}
\multiput(308.17,533.00)(8.000,23.990){2}{\rule{0.400pt}{0.725pt}}
\put(134.0,560.0){\rule[-0.200pt]{40.471pt}{0.400pt}}
\put(347,558.67){\rule{1.927pt}{0.400pt}}
\multiput(347.00,559.17)(4.000,-1.000){2}{\rule{0.964pt}{0.400pt}}
\put(355,557.67){\rule{1.686pt}{0.400pt}}
\multiput(355.00,558.17)(3.500,-1.000){2}{\rule{0.843pt}{0.400pt}}
\put(362,558.17){\rule{1.700pt}{0.400pt}}
\multiput(362.00,557.17)(4.472,2.000){2}{\rule{0.850pt}{0.400pt}}
\put(317.0,560.0){\rule[-0.200pt]{7.227pt}{0.400pt}}
\multiput(385.59,555.02)(0.488,-1.418){13}{\rule{0.117pt}{1.200pt}}
\multiput(384.17,557.51)(8.000,-19.509){2}{\rule{0.400pt}{0.600pt}}
\multiput(393.59,538.00)(0.485,1.637){11}{\rule{0.117pt}{1.357pt}}
\multiput(392.17,538.00)(7.000,19.183){2}{\rule{0.400pt}{0.679pt}}
\put(370.0,560.0){\rule[-0.200pt]{3.613pt}{0.400pt}}
\multiput(408.00,558.93)(0.581,-0.482){9}{\rule{0.567pt}{0.116pt}}
\multiput(408.00,559.17)(5.824,-6.000){2}{\rule{0.283pt}{0.400pt}}
\multiput(415.00,554.59)(0.671,0.482){9}{\rule{0.633pt}{0.116pt}}
\multiput(415.00,553.17)(6.685,6.000){2}{\rule{0.317pt}{0.400pt}}
\put(400.0,560.0){\rule[-0.200pt]{1.927pt}{0.400pt}}
\multiput(430.00,558.95)(1.355,-0.447){3}{\rule{1.033pt}{0.108pt}}
\multiput(430.00,559.17)(4.855,-3.000){2}{\rule{0.517pt}{0.400pt}}
\multiput(437.00,557.61)(1.579,0.447){3}{\rule{1.167pt}{0.108pt}}
\multiput(437.00,556.17)(5.579,3.000){2}{\rule{0.583pt}{0.400pt}}
\multiput(445.59,555.55)(0.485,-1.255){11}{\rule{0.117pt}{1.071pt}}
\multiput(444.17,557.78)(7.000,-14.776){2}{\rule{0.400pt}{0.536pt}}
\multiput(452.59,543.00)(0.488,1.088){13}{\rule{0.117pt}{0.950pt}}
\multiput(451.17,543.00)(8.000,15.028){2}{\rule{0.400pt}{0.475pt}}
\multiput(460.00,558.95)(1.355,-0.447){3}{\rule{1.033pt}{0.108pt}}
\multiput(460.00,559.17)(4.855,-3.000){2}{\rule{0.517pt}{0.400pt}}
\put(467,557.17){\rule{1.700pt}{0.400pt}}
\multiput(467.00,556.17)(4.472,2.000){2}{\rule{0.850pt}{0.400pt}}
\put(475,558.67){\rule{1.686pt}{0.400pt}}
\multiput(475.00,558.17)(3.500,1.000){2}{\rule{0.843pt}{0.400pt}}
\multiput(482.00,558.93)(0.710,-0.477){7}{\rule{0.660pt}{0.115pt}}
\multiput(482.00,559.17)(5.630,-5.000){2}{\rule{0.330pt}{0.400pt}}
\put(489,553.67){\rule{1.927pt}{0.400pt}}
\multiput(489.00,554.17)(4.000,-1.000){2}{\rule{0.964pt}{0.400pt}}
\multiput(497.00,554.59)(0.710,0.477){7}{\rule{0.660pt}{0.115pt}}
\multiput(497.00,553.17)(5.630,5.000){2}{\rule{0.330pt}{0.400pt}}
\put(504,558.67){\rule{1.927pt}{0.400pt}}
\multiput(504.00,558.17)(4.000,1.000){2}{\rule{0.964pt}{0.400pt}}
\put(423.0,560.0){\rule[-0.200pt]{1.686pt}{0.400pt}}
\put(520,558.67){\rule{1.686pt}{0.400pt}}
\multiput(520.00,559.17)(3.500,-1.000){2}{\rule{0.843pt}{0.400pt}}
\multiput(527.00,557.93)(0.569,-0.485){11}{\rule{0.557pt}{0.117pt}}
\multiput(527.00,558.17)(6.844,-7.000){2}{\rule{0.279pt}{0.400pt}}
\multiput(535.59,548.26)(0.485,-1.026){11}{\rule{0.117pt}{0.900pt}}
\multiput(534.17,550.13)(7.000,-12.132){2}{\rule{0.400pt}{0.450pt}}
\multiput(542.59,538.00)(0.488,1.484){13}{\rule{0.117pt}{1.250pt}}
\multiput(541.17,538.00)(8.000,20.406){2}{\rule{0.400pt}{0.625pt}}
\multiput(550.59,557.74)(0.485,-0.874){11}{\rule{0.117pt}{0.786pt}}
\multiput(549.17,559.37)(7.000,-10.369){2}{\rule{0.400pt}{0.393pt}}
\multiput(557.59,549.00)(0.488,0.758){13}{\rule{0.117pt}{0.700pt}}
\multiput(556.17,549.00)(8.000,10.547){2}{\rule{0.400pt}{0.350pt}}
\multiput(565.59,552.76)(0.485,-2.476){11}{\rule{0.117pt}{1.986pt}}
\multiput(564.17,556.88)(7.000,-28.879){2}{\rule{0.400pt}{0.993pt}}
\multiput(572.59,528.00)(0.485,1.255){11}{\rule{0.117pt}{1.071pt}}
\multiput(571.17,528.00)(7.000,14.776){2}{\rule{0.400pt}{0.536pt}}
\multiput(579.00,545.59)(0.494,0.488){13}{\rule{0.500pt}{0.117pt}}
\multiput(579.00,544.17)(6.962,8.000){2}{\rule{0.250pt}{0.400pt}}
\multiput(587.59,549.03)(0.485,-1.103){11}{\rule{0.117pt}{0.957pt}}
\multiput(586.17,551.01)(7.000,-13.013){2}{\rule{0.400pt}{0.479pt}}
\multiput(594.00,536.94)(1.066,-0.468){5}{\rule{0.900pt}{0.113pt}}
\multiput(594.00,537.17)(6.132,-4.000){2}{\rule{0.450pt}{0.400pt}}
\multiput(602.59,534.00)(0.485,2.323){11}{\rule{0.117pt}{1.871pt}}
\multiput(601.17,534.00)(7.000,27.116){2}{\rule{0.400pt}{0.936pt}}
\multiput(609.59,559.19)(0.488,-1.682){13}{\rule{0.117pt}{1.400pt}}
\multiput(608.17,562.09)(8.000,-23.094){2}{\rule{0.400pt}{0.700pt}}
\multiput(617.59,534.32)(0.485,-1.332){11}{\rule{0.117pt}{1.129pt}}
\multiput(616.17,536.66)(7.000,-15.658){2}{\rule{0.400pt}{0.564pt}}
\put(624,519.67){\rule{1.686pt}{0.400pt}}
\multiput(624.00,520.17)(3.500,-1.000){2}{\rule{0.843pt}{0.400pt}}
\multiput(631.59,520.00)(0.488,2.145){13}{\rule{0.117pt}{1.750pt}}
\multiput(630.17,520.00)(8.000,29.368){2}{\rule{0.400pt}{0.875pt}}
\multiput(639.59,550.21)(0.485,-0.721){11}{\rule{0.117pt}{0.671pt}}
\multiput(638.17,551.61)(7.000,-8.606){2}{\rule{0.400pt}{0.336pt}}
\multiput(646.59,543.00)(0.485,2.705){11}{\rule{0.117pt}{2.157pt}}
\multiput(645.17,543.00)(7.000,31.523){2}{\rule{0.400pt}{1.079pt}}
\multiput(653.59,573.60)(0.488,-1.550){13}{\rule{0.117pt}{1.300pt}}
\multiput(652.17,576.30)(8.000,-21.302){2}{\rule{0.400pt}{0.650pt}}
\multiput(661.59,551.50)(0.485,-0.950){11}{\rule{0.117pt}{0.843pt}}
\multiput(660.17,553.25)(7.000,-11.251){2}{\rule{0.400pt}{0.421pt}}
\multiput(668.59,536.40)(0.488,-1.616){13}{\rule{0.117pt}{1.350pt}}
\multiput(667.17,539.20)(8.000,-22.198){2}{\rule{0.400pt}{0.675pt}}
\multiput(676.59,517.00)(0.485,1.865){11}{\rule{0.117pt}{1.529pt}}
\multiput(675.17,517.00)(7.000,21.827){2}{\rule{0.400pt}{0.764pt}}
\multiput(683.59,536.19)(0.488,-1.682){13}{\rule{0.117pt}{1.400pt}}
\multiput(682.17,539.09)(8.000,-23.094){2}{\rule{0.400pt}{0.700pt}}
\put(512.0,560.0){\rule[-0.200pt]{1.927pt}{0.400pt}}
\multiput(698.59,516.00)(0.488,15.814){13}{\rule{0.117pt}{12.100pt}}
\multiput(697.17,516.00)(8.000,214.886){2}{\rule{0.400pt}{6.050pt}}
\multiput(706.59,749.89)(0.485,-1.789){11}{\rule{0.117pt}{1.471pt}}
\multiput(705.17,752.95)(7.000,-20.946){2}{\rule{0.400pt}{0.736pt}}
\multiput(713.59,680.35)(0.485,-16.435){11}{\rule{0.117pt}{12.443pt}}
\multiput(712.17,706.17)(7.000,-190.174){2}{\rule{0.400pt}{6.221pt}}
\multiput(720.59,516.00)(0.488,1.418){13}{\rule{0.117pt}{1.200pt}}
\multiput(719.17,516.00)(8.000,19.509){2}{\rule{0.400pt}{0.600pt}}
\multiput(728.59,534.68)(0.488,-0.890){13}{\rule{0.117pt}{0.800pt}}
\multiput(727.17,536.34)(8.000,-12.340){2}{\rule{0.400pt}{0.400pt}}
\put(736,522.17){\rule{1.500pt}{0.400pt}}
\multiput(736.00,523.17)(3.887,-2.000){2}{\rule{0.750pt}{0.400pt}}
\multiput(743.00,522.59)(0.581,0.482){9}{\rule{0.567pt}{0.116pt}}
\multiput(743.00,521.17)(5.824,6.000){2}{\rule{0.283pt}{0.400pt}}
\multiput(750.00,528.59)(0.821,0.477){7}{\rule{0.740pt}{0.115pt}}
\multiput(750.00,527.17)(6.464,5.000){2}{\rule{0.370pt}{0.400pt}}
\multiput(758.59,533.00)(0.485,2.018){11}{\rule{0.117pt}{1.643pt}}
\multiput(757.17,533.00)(7.000,23.590){2}{\rule{0.400pt}{0.821pt}}
\multiput(765.59,560.00)(0.485,0.569){11}{\rule{0.117pt}{0.557pt}}
\multiput(764.17,560.00)(7.000,6.844){2}{\rule{0.400pt}{0.279pt}}
\multiput(772.59,561.15)(0.488,-2.013){13}{\rule{0.117pt}{1.650pt}}
\multiput(771.17,564.58)(8.000,-27.575){2}{\rule{0.400pt}{0.825pt}}
\multiput(780.59,531.13)(0.485,-1.713){11}{\rule{0.117pt}{1.414pt}}
\multiput(779.17,534.06)(7.000,-20.065){2}{\rule{0.400pt}{0.707pt}}
\multiput(787.59,514.00)(0.488,0.560){13}{\rule{0.117pt}{0.550pt}}
\multiput(786.17,514.00)(8.000,7.858){2}{\rule{0.400pt}{0.275pt}}
\multiput(795.59,523.00)(0.485,2.933){11}{\rule{0.117pt}{2.329pt}}
\multiput(794.17,523.00)(7.000,34.167){2}{\rule{0.400pt}{1.164pt}}
\multiput(802.59,554.47)(0.485,-2.247){11}{\rule{0.117pt}{1.814pt}}
\multiput(801.17,558.23)(7.000,-26.234){2}{\rule{0.400pt}{0.907pt}}
\multiput(809.59,529.51)(0.488,-0.626){13}{\rule{0.117pt}{0.600pt}}
\multiput(808.17,530.75)(8.000,-8.755){2}{\rule{0.400pt}{0.300pt}}
\multiput(817.59,522.00)(0.485,8.731){11}{\rule{0.117pt}{6.671pt}}
\multiput(816.17,522.00)(7.000,101.153){2}{\rule{0.400pt}{3.336pt}}
\multiput(824.59,606.46)(0.485,-9.646){11}{\rule{0.117pt}{7.357pt}}
\multiput(823.17,621.73)(7.000,-111.730){2}{\rule{0.400pt}{3.679pt}}
\multiput(831.00,510.59)(0.494,0.488){13}{\rule{0.500pt}{0.117pt}}
\multiput(831.00,509.17)(6.962,8.000){2}{\rule{0.250pt}{0.400pt}}
\multiput(839.59,518.00)(0.485,0.645){11}{\rule{0.117pt}{0.614pt}}
\multiput(838.17,518.00)(7.000,7.725){2}{\rule{0.400pt}{0.307pt}}
\multiput(846.59,523.26)(0.485,-1.026){11}{\rule{0.117pt}{0.900pt}}
\multiput(845.17,525.13)(7.000,-12.132){2}{\rule{0.400pt}{0.450pt}}
\multiput(853.59,513.00)(0.488,12.512){13}{\rule{0.117pt}{9.600pt}}
\multiput(852.17,513.00)(8.000,170.075){2}{\rule{0.400pt}{4.800pt}}
\multiput(861.59,673.41)(0.485,-9.341){11}{\rule{0.117pt}{7.129pt}}
\multiput(860.17,688.20)(7.000,-108.204){2}{\rule{0.400pt}{3.564pt}}
\multiput(868.59,570.10)(0.485,-3.010){11}{\rule{0.117pt}{2.386pt}}
\multiput(867.17,575.05)(7.000,-35.048){2}{\rule{0.400pt}{1.193pt}}
\multiput(875.59,534.60)(0.488,-1.550){13}{\rule{0.117pt}{1.300pt}}
\multiput(874.17,537.30)(8.000,-21.302){2}{\rule{0.400pt}{0.650pt}}
\multiput(883.59,516.00)(0.485,2.018){11}{\rule{0.117pt}{1.643pt}}
\multiput(882.17,516.00)(7.000,23.590){2}{\rule{0.400pt}{0.821pt}}
\multiput(890.59,536.42)(0.485,-1.942){11}{\rule{0.117pt}{1.586pt}}
\multiput(889.17,539.71)(7.000,-22.709){2}{\rule{0.400pt}{0.793pt}}
\multiput(897.59,517.00)(0.488,1.088){13}{\rule{0.117pt}{0.950pt}}
\multiput(896.17,517.00)(8.000,15.028){2}{\rule{0.400pt}{0.475pt}}
\multiput(905.59,534.00)(0.485,12.316){11}{\rule{0.117pt}{9.357pt}}
\multiput(904.17,534.00)(7.000,142.579){2}{\rule{0.400pt}{4.679pt}}
\multiput(912.59,687.99)(0.485,-2.399){11}{\rule{0.117pt}{1.929pt}}
\multiput(911.17,692.00)(7.000,-27.997){2}{\rule{0.400pt}{0.964pt}}
\multiput(919.59,644.07)(0.488,-6.173){13}{\rule{0.117pt}{4.800pt}}
\multiput(918.17,654.04)(8.000,-84.037){2}{\rule{0.400pt}{2.400pt}}
\multiput(927.59,570.00)(0.485,1.103){11}{\rule{0.117pt}{0.957pt}}
\multiput(926.17,570.00)(7.000,13.013){2}{\rule{0.400pt}{0.479pt}}
\multiput(934.59,585.00)(0.485,5.374){11}{\rule{0.117pt}{4.157pt}}
\multiput(933.17,585.00)(7.000,62.372){2}{\rule{0.400pt}{2.079pt}}
\multiput(941.59,642.92)(0.488,-3.994){13}{\rule{0.117pt}{3.150pt}}
\multiput(940.17,649.46)(8.000,-54.462){2}{\rule{0.400pt}{1.575pt}}
\multiput(949.59,595.00)(0.485,2.933){11}{\rule{0.117pt}{2.329pt}}
\multiput(948.17,595.00)(7.000,34.167){2}{\rule{0.400pt}{1.164pt}}
\multiput(956.59,616.51)(0.485,-5.451){11}{\rule{0.117pt}{4.214pt}}
\multiput(955.17,625.25)(7.000,-63.253){2}{\rule{0.400pt}{2.107pt}}
\multiput(963.59,559.51)(0.488,-0.626){13}{\rule{0.117pt}{0.600pt}}
\multiput(962.17,560.75)(8.000,-8.755){2}{\rule{0.400pt}{0.300pt}}
\multiput(971.00,552.60)(0.920,0.468){5}{\rule{0.800pt}{0.113pt}}
\multiput(971.00,551.17)(5.340,4.000){2}{\rule{0.400pt}{0.400pt}}
\multiput(978.59,546.10)(0.485,-3.010){11}{\rule{0.117pt}{2.386pt}}
\multiput(977.17,551.05)(7.000,-35.048){2}{\rule{0.400pt}{1.193pt}}
\multiput(985.00,516.59)(0.494,0.488){13}{\rule{0.500pt}{0.117pt}}
\multiput(985.00,515.17)(6.962,8.000){2}{\rule{0.250pt}{0.400pt}}
\multiput(993.59,519.55)(0.485,-1.255){11}{\rule{0.117pt}{1.071pt}}
\multiput(992.17,521.78)(7.000,-14.776){2}{\rule{0.400pt}{0.536pt}}
\multiput(1000.59,503.26)(0.485,-1.026){11}{\rule{0.117pt}{0.900pt}}
\multiput(999.17,505.13)(7.000,-12.132){2}{\rule{0.400pt}{0.450pt}}
\put(1007,491.67){\rule{1.927pt}{0.400pt}}
\multiput(1007.00,492.17)(4.000,-1.000){2}{\rule{0.964pt}{0.400pt}}
\multiput(1015.59,487.08)(0.485,-1.408){11}{\rule{0.117pt}{1.186pt}}
\multiput(1014.17,489.54)(7.000,-16.539){2}{\rule{0.400pt}{0.593pt}}
\multiput(1022.59,431.55)(0.485,-13.155){11}{\rule{0.117pt}{9.986pt}}
\multiput(1021.17,452.27)(7.000,-152.274){2}{\rule{0.400pt}{4.993pt}}
\multiput(1029.59,300.00)(0.488,3.531){13}{\rule{0.117pt}{2.800pt}}
\multiput(1028.17,300.00)(8.000,48.188){2}{\rule{0.400pt}{1.400pt}}
\multiput(1037.59,299.50)(0.485,-17.350){11}{\rule{0.117pt}{13.129pt}}
\multiput(1036.17,326.75)(7.000,-200.751){2}{\rule{0.400pt}{6.564pt}}
\multiput(1044.59,126.00)(0.485,16.435){11}{\rule{0.117pt}{12.443pt}}
\multiput(1043.17,126.00)(7.000,190.174){2}{\rule{0.400pt}{6.221pt}}
\multiput(1051.59,342.00)(0.488,9.409){13}{\rule{0.117pt}{7.250pt}}
\multiput(1050.17,342.00)(8.000,127.952){2}{\rule{0.400pt}{3.625pt}}
\multiput(1059.59,485.00)(0.485,3.849){11}{\rule{0.117pt}{3.014pt}}
\multiput(1058.17,485.00)(7.000,44.744){2}{\rule{0.400pt}{1.507pt}}
\multiput(1066.59,471.78)(0.485,-20.478){11}{\rule{0.117pt}{15.471pt}}
\multiput(1065.17,503.89)(7.000,-236.888){2}{\rule{0.400pt}{7.736pt}}
\multiput(1073.59,267.00)(0.488,8.616){13}{\rule{0.117pt}{6.650pt}}
\multiput(1072.17,267.00)(8.000,117.198){2}{\rule{0.400pt}{3.325pt}}
\multiput(1081.59,398.00)(0.485,15.138){11}{\rule{0.117pt}{11.471pt}}
\multiput(1080.17,398.00)(7.000,175.190){2}{\rule{0.400pt}{5.736pt}}
\multiput(1088.59,597.00)(0.485,1.789){11}{\rule{0.117pt}{1.471pt}}
\multiput(1087.17,597.00)(7.000,20.946){2}{\rule{0.400pt}{0.736pt}}
\multiput(1095.00,621.59)(0.821,0.477){7}{\rule{0.740pt}{0.115pt}}
\multiput(1095.00,620.17)(6.464,5.000){2}{\rule{0.370pt}{0.400pt}}
\multiput(1103.59,626.00)(0.485,2.933){11}{\rule{0.117pt}{2.329pt}}
\multiput(1102.17,626.00)(7.000,34.167){2}{\rule{0.400pt}{1.164pt}}
\multiput(1110.59,665.00)(0.485,4.383){11}{\rule{0.117pt}{3.414pt}}
\multiput(1109.17,665.00)(7.000,50.913){2}{\rule{0.400pt}{1.707pt}}
\multiput(1117.59,713.04)(0.488,-3.003){13}{\rule{0.117pt}{2.400pt}}
\multiput(1116.17,718.02)(8.000,-41.019){2}{\rule{0.400pt}{1.200pt}}
\multiput(1125.59,677.00)(0.485,3.239){11}{\rule{0.117pt}{2.557pt}}
\multiput(1124.17,677.00)(7.000,37.693){2}{\rule{0.400pt}{1.279pt}}
\multiput(1132.59,715.32)(0.485,-1.332){11}{\rule{0.117pt}{1.129pt}}
\multiput(1131.17,717.66)(7.000,-15.658){2}{\rule{0.400pt}{0.564pt}}
\multiput(1139.59,702.00)(0.488,5.645){13}{\rule{0.117pt}{4.400pt}}
\multiput(1138.17,702.00)(8.000,76.868){2}{\rule{0.400pt}{2.200pt}}
\multiput(1147.59,775.49)(0.485,-3.849){11}{\rule{0.117pt}{3.014pt}}
\multiput(1146.17,781.74)(7.000,-44.744){2}{\rule{0.400pt}{1.507pt}}
\multiput(1154.59,737.00)(0.485,8.578){11}{\rule{0.117pt}{6.557pt}}
\multiput(1153.17,737.00)(7.000,99.390){2}{\rule{0.400pt}{3.279pt}}
\multiput(1161.59,828.41)(0.488,-6.701){13}{\rule{0.117pt}{5.200pt}}
\multiput(1160.17,839.21)(8.000,-91.207){2}{\rule{0.400pt}{2.600pt}}
\multiput(1169.00,746.93)(0.492,-0.485){11}{\rule{0.500pt}{0.117pt}}
\multiput(1169.00,747.17)(5.962,-7.000){2}{\rule{0.250pt}{0.400pt}}
\multiput(1176.00,739.94)(0.920,-0.468){5}{\rule{0.800pt}{0.113pt}}
\multiput(1176.00,740.17)(5.340,-4.000){2}{\rule{0.400pt}{0.400pt}}
\multiput(1183.59,737.00)(0.488,7.824){13}{\rule{0.117pt}{6.050pt}}
\multiput(1182.17,737.00)(8.000,106.443){2}{\rule{0.400pt}{3.025pt}}
\multiput(1191.59,826.65)(0.485,-9.265){11}{\rule{0.117pt}{7.071pt}}
\multiput(1190.17,841.32)(7.000,-107.323){2}{\rule{0.400pt}{3.536pt}}
\multiput(1198.59,734.00)(0.485,6.442){11}{\rule{0.117pt}{4.957pt}}
\multiput(1197.17,734.00)(7.000,74.711){2}{\rule{0.400pt}{2.479pt}}
\multiput(1205.59,803.02)(0.488,-4.918){13}{\rule{0.117pt}{3.850pt}}
\multiput(1204.17,811.01)(8.000,-67.009){2}{\rule{0.400pt}{1.925pt}}
\multiput(1213.00,744.59)(0.581,0.482){9}{\rule{0.567pt}{0.116pt}}
\multiput(1213.00,743.17)(5.824,6.000){2}{\rule{0.283pt}{0.400pt}}
\multiput(1220.59,750.00)(0.485,1.103){11}{\rule{0.117pt}{0.957pt}}
\multiput(1219.17,750.00)(7.000,13.013){2}{\rule{0.400pt}{0.479pt}}
\multiput(1227.00,765.59)(0.494,0.488){13}{\rule{0.500pt}{0.117pt}}
\multiput(1227.00,764.17)(6.962,8.000){2}{\rule{0.250pt}{0.400pt}}
\multiput(1235.59,773.00)(0.485,1.255){11}{\rule{0.117pt}{1.071pt}}
\multiput(1234.17,773.00)(7.000,14.776){2}{\rule{0.400pt}{0.536pt}}
\multiput(1242.59,790.00)(0.485,0.569){11}{\rule{0.117pt}{0.557pt}}
\multiput(1241.17,790.00)(7.000,6.844){2}{\rule{0.400pt}{0.279pt}}
\multiput(1249.59,794.68)(0.488,-0.890){13}{\rule{0.117pt}{0.800pt}}
\multiput(1248.17,796.34)(8.000,-12.340){2}{\rule{0.400pt}{0.400pt}}
\put(1257,782.17){\rule{1.500pt}{0.400pt}}
\multiput(1257.00,783.17)(3.887,-2.000){2}{\rule{0.750pt}{0.400pt}}
\multiput(1264.59,782.00)(0.485,1.484){11}{\rule{0.117pt}{1.243pt}}
\multiput(1263.17,782.00)(7.000,17.420){2}{\rule{0.400pt}{0.621pt}}
\multiput(1271.59,797.64)(0.488,-1.220){13}{\rule{0.117pt}{1.050pt}}
\multiput(1270.17,799.82)(8.000,-16.821){2}{\rule{0.400pt}{0.525pt}}
\multiput(1279.59,783.00)(0.485,2.018){11}{\rule{0.117pt}{1.643pt}}
\multiput(1278.17,783.00)(7.000,23.590){2}{\rule{0.400pt}{0.821pt}}
\multiput(1286.59,804.37)(0.485,-1.637){11}{\rule{0.117pt}{1.357pt}}
\multiput(1285.17,807.18)(7.000,-19.183){2}{\rule{0.400pt}{0.679pt}}
\multiput(1293.59,788.00)(0.488,1.220){13}{\rule{0.117pt}{1.050pt}}
\multiput(1292.17,788.00)(8.000,16.821){2}{\rule{0.400pt}{0.525pt}}
\multiput(1301.00,805.93)(0.492,-0.485){11}{\rule{0.500pt}{0.117pt}}
\multiput(1301.00,806.17)(5.962,-7.000){2}{\rule{0.250pt}{0.400pt}}
\multiput(1308.59,800.00)(0.485,2.323){11}{\rule{0.117pt}{1.871pt}}
\multiput(1307.17,800.00)(7.000,27.116){2}{\rule{0.400pt}{0.936pt}}
\multiput(1315.59,818.96)(0.485,-3.696){11}{\rule{0.117pt}{2.900pt}}
\multiput(1314.17,824.98)(7.000,-42.981){2}{\rule{0.400pt}{1.450pt}}
\multiput(1322.59,782.00)(0.488,3.333){13}{\rule{0.117pt}{2.650pt}}
\multiput(1321.17,782.00)(8.000,45.500){2}{\rule{0.400pt}{1.325pt}}
\multiput(1330.59,825.71)(0.485,-2.171){11}{\rule{0.117pt}{1.757pt}}
\multiput(1329.17,829.35)(7.000,-25.353){2}{\rule{0.400pt}{0.879pt}}
\multiput(1337.59,804.00)(0.485,3.086){11}{\rule{0.117pt}{2.443pt}}
\multiput(1336.17,804.00)(7.000,35.930){2}{\rule{0.400pt}{1.221pt}}
\multiput(1344.59,832.55)(0.488,-3.796){13}{\rule{0.117pt}{3.000pt}}
\multiput(1343.17,838.77)(8.000,-51.773){2}{\rule{0.400pt}{1.500pt}}
\multiput(1352.59,787.00)(0.485,1.255){11}{\rule{0.117pt}{1.071pt}}
\multiput(1351.17,787.00)(7.000,14.776){2}{\rule{0.400pt}{0.536pt}}
\put(691.0,516.0){\rule[-0.200pt]{1.686pt}{0.400pt}}
\end{picture}

%% file: FIGPLBN3.tex
\setlength{\unitlength}{0.240900pt}
\ifx\plotpoint\undefined\newsavebox{\plotpoint}\fi
\begin{picture}(1500,900)(0,0)
\font\gnuplot=cmr10 at 10pt
\gnuplot
\sbox{\plotpoint}{\rule[-0.200pt]{0.400pt}{0.400pt}}%
\put(120.0,123.0){\rule[-0.200pt]{4.818pt}{0.400pt}}
\put(100,123){\makebox(0,0)[r]{-60}}
\put(1419.0,123.0){\rule[-0.200pt]{4.818pt}{0.400pt}}
\put(120.0,246.0){\rule[-0.200pt]{4.818pt}{0.400pt}}
\put(100,246){\makebox(0,0)[r]{-50}}
\put(1419.0,246.0){\rule[-0.200pt]{4.818pt}{0.400pt}}
\put(120.0,369.0){\rule[-0.200pt]{4.818pt}{0.400pt}}
\put(100,369){\makebox(0,0)[r]{-40}}
\put(1419.0,369.0){\rule[-0.200pt]{4.818pt}{0.400pt}}
\put(120.0,492.0){\rule[-0.200pt]{4.818pt}{0.400pt}}
\put(100,492){\makebox(0,0)[r]{-30}}
\put(1419.0,492.0){\rule[-0.200pt]{4.818pt}{0.400pt}}
\put(120.0,614.0){\rule[-0.200pt]{4.818pt}{0.400pt}}
\put(100,614){\makebox(0,0)[r]{-20}}
\put(1419.0,614.0){\rule[-0.200pt]{4.818pt}{0.400pt}}
\put(120.0,737.0){\rule[-0.200pt]{4.818pt}{0.400pt}}
\put(100,737){\makebox(0,0)[r]{-10}}
\put(1419.0,737.0){\rule[-0.200pt]{4.818pt}{0.400pt}}
\put(120.0,860.0){\rule[-0.200pt]{4.818pt}{0.400pt}}
\put(100,860){\makebox(0,0)[r]{0}}
\put(1419.0,860.0){\rule[-0.200pt]{4.818pt}{0.400pt}}
\put(120.0,123.0){\rule[-0.200pt]{0.400pt}{4.818pt}}
\put(120,82){\makebox(0,0){0}}
\put(120.0,840.0){\rule[-0.200pt]{0.400pt}{4.818pt}}
\put(267.0,123.0){\rule[-0.200pt]{0.400pt}{4.818pt}}
\put(267,82){\makebox(0,0){1}}
\put(267.0,840.0){\rule[-0.200pt]{0.400pt}{4.818pt}}
\put(413.0,123.0){\rule[-0.200pt]{0.400pt}{4.818pt}}
\put(413,82){\makebox(0,0){2}}
\put(413.0,840.0){\rule[-0.200pt]{0.400pt}{4.818pt}}
\put(560.0,123.0){\rule[-0.200pt]{0.400pt}{4.818pt}}
\put(560,82){\makebox(0,0){3}}
\put(560.0,840.0){\rule[-0.200pt]{0.400pt}{4.818pt}}
\put(706.0,123.0){\rule[-0.200pt]{0.400pt}{4.818pt}}
\put(706,82){\makebox(0,0){4}}
\put(706.0,840.0){\rule[-0.200pt]{0.400pt}{4.818pt}}
\put(853.0,123.0){\rule[-0.200pt]{0.400pt}{4.818pt}}
\put(853,82){\makebox(0,0){5}}
\put(853.0,840.0){\rule[-0.200pt]{0.400pt}{4.818pt}}
\put(999.0,123.0){\rule[-0.200pt]{0.400pt}{4.818pt}}
\put(999,82){\makebox(0,0){6}}
\put(999.0,840.0){\rule[-0.200pt]{0.400pt}{4.818pt}}
\put(1146.0,123.0){\rule[-0.200pt]{0.400pt}{4.818pt}}
\put(1146,82){\makebox(0,0){7}}
\put(1146.0,840.0){\rule[-0.200pt]{0.400pt}{4.818pt}}
\put(1292.0,123.0){\rule[-0.200pt]{0.400pt}{4.818pt}}
\put(1292,82){\makebox(0,0){8}}
\put(1292.0,840.0){\rule[-0.200pt]{0.400pt}{4.818pt}}
\put(1439.0,123.0){\rule[-0.200pt]{0.400pt}{4.818pt}}
\put(1439,82){\makebox(0,0){9}}
\put(1439.0,840.0){\rule[-0.200pt]{0.400pt}{4.818pt}}
\put(120.0,123.0){\rule[-0.200pt]{317.747pt}{0.400pt}}
\put(1439.0,123.0){\rule[-0.200pt]{0.400pt}{177.543pt}}
\put(120.0,860.0){\rule[-0.200pt]{317.747pt}{0.400pt}}
\put(779,21){\makebox(0,0){n(efolds)}}
\put(120.0,123.0){\rule[-0.200pt]{0.400pt}{177.543pt}}
\put(134,178){\usebox{\plotpoint}}
\multiput(134.00,178.59)(1.267,0.477){7}{\rule{1.060pt}{0.115pt}}
\multiput(134.00,177.17)(9.800,5.000){2}{\rule{0.530pt}{0.400pt}}
\put(172,182.67){\rule{3.132pt}{0.400pt}}
\multiput(172.00,182.17)(6.500,1.000){2}{\rule{1.566pt}{0.400pt}}
\put(185,184.17){\rule{2.900pt}{0.400pt}}
\multiput(185.00,183.17)(7.981,2.000){2}{\rule{1.450pt}{0.400pt}}
\put(199,184.17){\rule{2.900pt}{0.400pt}}
\multiput(199.00,185.17)(7.981,-2.000){2}{\rule{1.450pt}{0.400pt}}
\multiput(213.00,182.95)(3.141,-0.447){3}{\rule{2.100pt}{0.108pt}}
\multiput(213.00,183.17)(10.641,-3.000){2}{\rule{1.050pt}{0.400pt}}
\multiput(228.00,179.92)(0.779,-0.492){19}{\rule{0.718pt}{0.118pt}}
\multiput(228.00,180.17)(15.509,-11.000){2}{\rule{0.359pt}{0.400pt}}
\put(245,170.17){\rule{4.900pt}{0.400pt}}
\multiput(245.00,169.17)(13.830,2.000){2}{\rule{2.450pt}{0.400pt}}
\multiput(269.00,170.93)(0.798,-0.485){11}{\rule{0.729pt}{0.117pt}}
\multiput(269.00,171.17)(9.488,-7.000){2}{\rule{0.364pt}{0.400pt}}
\multiput(280.00,165.59)(0.581,0.482){9}{\rule{0.567pt}{0.116pt}}
\multiput(280.00,164.17)(5.824,6.000){2}{\rule{0.283pt}{0.400pt}}
\multiput(287.59,171.00)(0.485,0.645){11}{\rule{0.117pt}{0.614pt}}
\multiput(286.17,171.00)(7.000,7.725){2}{\rule{0.400pt}{0.307pt}}
\multiput(294.00,180.59)(0.671,0.482){9}{\rule{0.633pt}{0.116pt}}
\multiput(294.00,179.17)(6.685,6.000){2}{\rule{0.317pt}{0.400pt}}
\put(302,185.67){\rule{1.686pt}{0.400pt}}
\multiput(302.00,185.17)(3.500,1.000){2}{\rule{0.843pt}{0.400pt}}
\put(309,185.67){\rule{1.686pt}{0.400pt}}
\multiput(309.00,186.17)(3.500,-1.000){2}{\rule{0.843pt}{0.400pt}}
\put(316,186.17){\rule{1.700pt}{0.400pt}}
\multiput(316.00,185.17)(4.472,2.000){2}{\rule{0.850pt}{0.400pt}}
\multiput(324.00,186.93)(0.581,-0.482){9}{\rule{0.567pt}{0.116pt}}
\multiput(324.00,187.17)(5.824,-6.000){2}{\rule{0.283pt}{0.400pt}}
\multiput(331.00,180.93)(0.495,-0.489){15}{\rule{0.500pt}{0.118pt}}
\multiput(331.00,181.17)(7.962,-9.000){2}{\rule{0.250pt}{0.400pt}}
\multiput(340.00,171.93)(0.710,-0.477){7}{\rule{0.660pt}{0.115pt}}
\multiput(340.00,172.17)(5.630,-5.000){2}{\rule{0.330pt}{0.400pt}}
\multiput(347.59,168.00)(0.485,1.026){11}{\rule{0.117pt}{0.900pt}}
\multiput(346.17,168.00)(7.000,12.132){2}{\rule{0.400pt}{0.450pt}}
\multiput(354.59,170.17)(0.488,-3.597){13}{\rule{0.117pt}{2.850pt}}
\multiput(353.17,176.08)(8.000,-49.085){2}{\rule{0.400pt}{1.425pt}}
\multiput(362.59,127.00)(0.485,3.772){11}{\rule{0.117pt}{2.957pt}}
\multiput(361.17,127.00)(7.000,43.862){2}{\rule{0.400pt}{1.479pt}}
\multiput(369.59,167.91)(0.489,-2.708){15}{\rule{0.118pt}{2.189pt}}
\multiput(368.17,172.46)(9.000,-42.457){2}{\rule{0.400pt}{1.094pt}}
\multiput(378.59,130.00)(0.485,3.772){11}{\rule{0.117pt}{2.957pt}}
\multiput(377.17,130.00)(7.000,43.862){2}{\rule{0.400pt}{1.479pt}}
\put(385,180.17){\rule{1.700pt}{0.400pt}}
\multiput(385.00,179.17)(4.472,2.000){2}{\rule{0.850pt}{0.400pt}}
\multiput(393.00,180.94)(0.920,-0.468){5}{\rule{0.800pt}{0.113pt}}
\multiput(393.00,181.17)(5.340,-4.000){2}{\rule{0.400pt}{0.400pt}}
\multiput(400.00,176.93)(0.581,-0.482){9}{\rule{0.567pt}{0.116pt}}
\multiput(400.00,177.17)(5.824,-6.000){2}{\rule{0.283pt}{0.400pt}}
\multiput(407.00,172.59)(0.671,0.482){9}{\rule{0.633pt}{0.116pt}}
\multiput(407.00,171.17)(6.685,6.000){2}{\rule{0.317pt}{0.400pt}}
\multiput(415.00,178.59)(0.710,0.477){7}{\rule{0.660pt}{0.115pt}}
\multiput(415.00,177.17)(5.630,5.000){2}{\rule{0.330pt}{0.400pt}}
\multiput(422.59,179.82)(0.489,-0.844){15}{\rule{0.118pt}{0.767pt}}
\multiput(421.17,181.41)(9.000,-13.409){2}{\rule{0.400pt}{0.383pt}}
\multiput(431.59,168.00)(0.485,1.103){11}{\rule{0.117pt}{0.957pt}}
\multiput(430.17,168.00)(7.000,13.013){2}{\rule{0.400pt}{0.479pt}}
\multiput(438.59,178.32)(0.485,-1.332){11}{\rule{0.117pt}{1.129pt}}
\multiput(437.17,180.66)(7.000,-15.658){2}{\rule{0.400pt}{0.564pt}}
\multiput(445.59,165.00)(0.488,1.088){13}{\rule{0.117pt}{0.950pt}}
\multiput(444.17,165.00)(8.000,15.028){2}{\rule{0.400pt}{0.475pt}}
\multiput(453.59,179.21)(0.485,-0.721){11}{\rule{0.117pt}{0.671pt}}
\multiput(452.17,180.61)(7.000,-8.606){2}{\rule{0.400pt}{0.336pt}}
\multiput(460.59,172.00)(0.485,0.721){11}{\rule{0.117pt}{0.671pt}}
\multiput(459.17,172.00)(7.000,8.606){2}{\rule{0.400pt}{0.336pt}}
\put(467,180.17){\rule{1.700pt}{0.400pt}}
\multiput(467.00,181.17)(4.472,-2.000){2}{\rule{0.850pt}{0.400pt}}
\multiput(475.59,172.23)(0.485,-2.323){11}{\rule{0.117pt}{1.871pt}}
\multiput(474.17,176.12)(7.000,-27.116){2}{\rule{0.400pt}{0.936pt}}
\multiput(482.59,149.00)(0.485,2.399){11}{\rule{0.117pt}{1.929pt}}
\multiput(481.17,149.00)(7.000,27.997){2}{\rule{0.400pt}{0.964pt}}
\put(146.0,183.0){\rule[-0.200pt]{6.263pt}{0.400pt}}
\multiput(497.00,179.93)(0.710,-0.477){7}{\rule{0.660pt}{0.115pt}}
\multiput(497.00,180.17)(5.630,-5.000){2}{\rule{0.330pt}{0.400pt}}
\multiput(504.59,170.97)(0.489,-1.427){15}{\rule{0.118pt}{1.211pt}}
\multiput(503.17,173.49)(9.000,-22.486){2}{\rule{0.400pt}{0.606pt}}
\multiput(513.59,151.00)(0.485,0.721){11}{\rule{0.117pt}{0.671pt}}
\multiput(512.17,151.00)(7.000,8.606){2}{\rule{0.400pt}{0.336pt}}
\put(489.0,181.0){\rule[-0.200pt]{1.927pt}{0.400pt}}
\multiput(527.59,161.00)(0.488,1.088){13}{\rule{0.117pt}{0.950pt}}
\multiput(526.17,161.00)(8.000,15.028){2}{\rule{0.400pt}{0.475pt}}
\multiput(535.00,178.60)(0.920,0.468){5}{\rule{0.800pt}{0.113pt}}
\multiput(535.00,177.17)(5.340,4.000){2}{\rule{0.400pt}{0.400pt}}
\multiput(542.00,180.94)(0.920,-0.468){5}{\rule{0.800pt}{0.113pt}}
\multiput(542.00,181.17)(5.340,-4.000){2}{\rule{0.400pt}{0.400pt}}
\put(549,176.67){\rule{1.927pt}{0.400pt}}
\multiput(549.00,177.17)(4.000,-1.000){2}{\rule{0.964pt}{0.400pt}}
\put(557,176.67){\rule{1.686pt}{0.400pt}}
\multiput(557.00,176.17)(3.500,1.000){2}{\rule{0.843pt}{0.400pt}}
\multiput(564.00,178.61)(1.355,0.447){3}{\rule{1.033pt}{0.108pt}}
\multiput(564.00,177.17)(4.855,3.000){2}{\rule{0.517pt}{0.400pt}}
\put(571,179.67){\rule{2.168pt}{0.400pt}}
\multiput(571.00,180.17)(4.500,-1.000){2}{\rule{1.084pt}{0.400pt}}
\multiput(580.59,177.51)(0.488,-0.626){13}{\rule{0.117pt}{0.600pt}}
\multiput(579.17,178.75)(8.000,-8.755){2}{\rule{0.400pt}{0.300pt}}
\multiput(588.59,170.00)(0.485,0.874){11}{\rule{0.117pt}{0.786pt}}
\multiput(587.17,170.00)(7.000,10.369){2}{\rule{0.400pt}{0.393pt}}
\multiput(595.59,176.60)(0.485,-1.560){11}{\rule{0.117pt}{1.300pt}}
\multiput(594.17,179.30)(7.000,-18.302){2}{\rule{0.400pt}{0.650pt}}
\multiput(602.59,161.00)(0.485,1.255){11}{\rule{0.117pt}{1.071pt}}
\multiput(601.17,161.00)(7.000,14.776){2}{\rule{0.400pt}{0.536pt}}
\multiput(609.00,178.60)(1.066,0.468){5}{\rule{0.900pt}{0.113pt}}
\multiput(609.00,177.17)(6.132,4.000){2}{\rule{0.450pt}{0.400pt}}
\put(617,180.17){\rule{1.500pt}{0.400pt}}
\multiput(617.00,181.17)(3.887,-2.000){2}{\rule{0.750pt}{0.400pt}}
\put(520.0,161.0){\rule[-0.200pt]{1.686pt}{0.400pt}}
\put(631,179.67){\rule{1.927pt}{0.400pt}}
\multiput(631.00,179.17)(4.000,1.000){2}{\rule{0.964pt}{0.400pt}}
\multiput(639.00,181.61)(1.355,0.447){3}{\rule{1.033pt}{0.108pt}}
\multiput(639.00,180.17)(4.855,3.000){2}{\rule{0.517pt}{0.400pt}}
\multiput(646.59,175.99)(0.485,-2.399){11}{\rule{0.117pt}{1.929pt}}
\multiput(645.17,180.00)(7.000,-27.997){2}{\rule{0.400pt}{0.964pt}}
\multiput(653.59,152.00)(0.488,1.947){13}{\rule{0.117pt}{1.600pt}}
\multiput(652.17,152.00)(8.000,26.679){2}{\rule{0.400pt}{0.800pt}}
\put(624.0,180.0){\rule[-0.200pt]{1.686pt}{0.400pt}}
\put(668,180.67){\rule{1.686pt}{0.400pt}}
\multiput(668.00,181.17)(3.500,-1.000){2}{\rule{0.843pt}{0.400pt}}
\put(661.0,182.0){\rule[-0.200pt]{1.686pt}{0.400pt}}
\multiput(683.59,177.63)(0.489,-0.902){15}{\rule{0.118pt}{0.811pt}}
\multiput(682.17,179.32)(9.000,-14.316){2}{\rule{0.400pt}{0.406pt}}
\put(692,164.67){\rule{1.686pt}{0.400pt}}
\multiput(692.00,164.17)(3.500,1.000){2}{\rule{0.843pt}{0.400pt}}
\multiput(699.59,166.00)(0.485,0.874){11}{\rule{0.117pt}{0.786pt}}
\multiput(698.17,166.00)(7.000,10.369){2}{\rule{0.400pt}{0.393pt}}
\multiput(706.00,176.95)(1.579,-0.447){3}{\rule{1.167pt}{0.108pt}}
\multiput(706.00,177.17)(5.579,-3.000){2}{\rule{0.583pt}{0.400pt}}
\multiput(714.59,172.69)(0.485,-0.569){11}{\rule{0.117pt}{0.557pt}}
\multiput(713.17,173.84)(7.000,-6.844){2}{\rule{0.400pt}{0.279pt}}
\multiput(721.59,167.00)(0.485,0.569){11}{\rule{0.117pt}{0.557pt}}
\multiput(720.17,167.00)(7.000,6.844){2}{\rule{0.400pt}{0.279pt}}
\multiput(728.00,173.93)(0.821,-0.477){7}{\rule{0.740pt}{0.115pt}}
\multiput(728.00,174.17)(6.464,-5.000){2}{\rule{0.370pt}{0.400pt}}
\multiput(736.59,166.26)(0.485,-1.026){11}{\rule{0.117pt}{0.900pt}}
\multiput(735.17,168.13)(7.000,-12.132){2}{\rule{0.400pt}{0.450pt}}
\multiput(743.59,156.00)(0.485,0.874){11}{\rule{0.117pt}{0.786pt}}
\multiput(742.17,156.00)(7.000,10.369){2}{\rule{0.400pt}{0.393pt}}
\put(750,166.67){\rule{1.927pt}{0.400pt}}
\multiput(750.00,167.17)(4.000,-1.000){2}{\rule{0.964pt}{0.400pt}}
\put(675.0,181.0){\rule[-0.200pt]{1.927pt}{0.400pt}}
\put(765,165.67){\rule{1.686pt}{0.400pt}}
\multiput(765.00,166.17)(3.500,-1.000){2}{\rule{0.843pt}{0.400pt}}
\put(772,166.17){\rule{1.700pt}{0.400pt}}
\multiput(772.00,165.17)(4.472,2.000){2}{\rule{0.850pt}{0.400pt}}
\multiput(780.59,168.00)(0.485,2.171){11}{\rule{0.117pt}{1.757pt}}
\multiput(779.17,168.00)(7.000,25.353){2}{\rule{0.400pt}{0.879pt}}
\multiput(787.00,197.59)(0.492,0.485){11}{\rule{0.500pt}{0.117pt}}
\multiput(787.00,196.17)(5.962,7.000){2}{\rule{0.250pt}{0.400pt}}
\put(794,202.67){\rule{1.686pt}{0.400pt}}
\multiput(794.00,203.17)(3.500,-1.000){2}{\rule{0.843pt}{0.400pt}}
\multiput(801.59,200.72)(0.488,-0.560){13}{\rule{0.117pt}{0.550pt}}
\multiput(800.17,201.86)(8.000,-7.858){2}{\rule{0.400pt}{0.275pt}}
\multiput(809.59,189.08)(0.485,-1.408){11}{\rule{0.117pt}{1.186pt}}
\multiput(808.17,191.54)(7.000,-16.539){2}{\rule{0.400pt}{0.593pt}}
\multiput(816.59,175.00)(0.485,1.408){11}{\rule{0.117pt}{1.186pt}}
\multiput(815.17,175.00)(7.000,16.539){2}{\rule{0.400pt}{0.593pt}}
\multiput(823.59,194.00)(0.488,3.135){13}{\rule{0.117pt}{2.500pt}}
\multiput(822.17,194.00)(8.000,42.811){2}{\rule{0.400pt}{1.250pt}}
\multiput(831.59,242.00)(0.485,11.858){11}{\rule{0.117pt}{9.014pt}}
\multiput(830.17,242.00)(7.000,137.290){2}{\rule{0.400pt}{4.507pt}}
\multiput(838.59,398.00)(0.485,1.255){11}{\rule{0.117pt}{1.071pt}}
\multiput(837.17,398.00)(7.000,14.776){2}{\rule{0.400pt}{0.536pt}}
\put(758.0,167.0){\rule[-0.200pt]{1.686pt}{0.400pt}}
\multiput(853.59,415.00)(0.485,6.061){11}{\rule{0.117pt}{4.671pt}}
\multiput(852.17,415.00)(7.000,70.304){2}{\rule{0.400pt}{2.336pt}}
\multiput(860.59,495.00)(0.485,15.977){11}{\rule{0.117pt}{12.100pt}}
\multiput(859.17,495.00)(7.000,184.886){2}{\rule{0.400pt}{6.050pt}}
\multiput(867.59,705.00)(0.488,0.692){13}{\rule{0.117pt}{0.650pt}}
\multiput(866.17,705.00)(8.000,9.651){2}{\rule{0.400pt}{0.325pt}}
\put(875,714.17){\rule{1.500pt}{0.400pt}}
\multiput(875.00,715.17)(3.887,-2.000){2}{\rule{0.750pt}{0.400pt}}
\multiput(882.59,710.03)(0.485,-1.103){11}{\rule{0.117pt}{0.957pt}}
\multiput(881.17,712.01)(7.000,-13.013){2}{\rule{0.400pt}{0.479pt}}
\multiput(889.59,699.00)(0.488,0.626){13}{\rule{0.117pt}{0.600pt}}
\multiput(888.17,699.00)(8.000,8.755){2}{\rule{0.400pt}{0.300pt}}
\multiput(897.59,709.00)(0.485,1.637){11}{\rule{0.117pt}{1.357pt}}
\multiput(896.17,709.00)(7.000,19.183){2}{\rule{0.400pt}{0.679pt}}
\multiput(904.59,731.00)(0.485,0.950){11}{\rule{0.117pt}{0.843pt}}
\multiput(903.17,731.00)(7.000,11.251){2}{\rule{0.400pt}{0.421pt}}
\put(845.0,415.0){\rule[-0.200pt]{1.927pt}{0.400pt}}
\multiput(919.59,744.00)(0.485,0.798){11}{\rule{0.117pt}{0.729pt}}
\multiput(918.17,744.00)(7.000,9.488){2}{\rule{0.400pt}{0.364pt}}
\put(926,753.17){\rule{1.500pt}{0.400pt}}
\multiput(926.00,754.17)(3.887,-2.000){2}{\rule{0.750pt}{0.400pt}}
\multiput(933.00,751.95)(1.579,-0.447){3}{\rule{1.167pt}{0.108pt}}
\multiput(933.00,752.17)(5.579,-3.000){2}{\rule{0.583pt}{0.400pt}}
\put(941,750.17){\rule{1.500pt}{0.400pt}}
\multiput(941.00,749.17)(3.887,2.000){2}{\rule{0.750pt}{0.400pt}}
\multiput(948.59,748.26)(0.485,-1.026){11}{\rule{0.117pt}{0.900pt}}
\multiput(947.17,750.13)(7.000,-12.132){2}{\rule{0.400pt}{0.450pt}}
\multiput(955.00,738.59)(0.569,0.485){11}{\rule{0.557pt}{0.117pt}}
\multiput(955.00,737.17)(6.844,7.000){2}{\rule{0.279pt}{0.400pt}}
\multiput(963.00,743.93)(0.581,-0.482){9}{\rule{0.567pt}{0.116pt}}
\multiput(963.00,744.17)(5.824,-6.000){2}{\rule{0.283pt}{0.400pt}}
\put(970,738.67){\rule{1.686pt}{0.400pt}}
\multiput(970.00,738.17)(3.500,1.000){2}{\rule{0.843pt}{0.400pt}}
\multiput(977.00,738.93)(0.821,-0.477){7}{\rule{0.740pt}{0.115pt}}
\multiput(977.00,739.17)(6.464,-5.000){2}{\rule{0.370pt}{0.400pt}}
\multiput(985.00,735.59)(0.821,0.477){7}{\rule{0.740pt}{0.115pt}}
\multiput(985.00,734.17)(6.464,5.000){2}{\rule{0.370pt}{0.400pt}}
\put(993,738.67){\rule{1.927pt}{0.400pt}}
\multiput(993.00,739.17)(4.000,-1.000){2}{\rule{0.964pt}{0.400pt}}
\multiput(1001.59,736.69)(0.485,-0.569){11}{\rule{0.117pt}{0.557pt}}
\multiput(1000.17,737.84)(7.000,-6.844){2}{\rule{0.400pt}{0.279pt}}
\multiput(1008.00,731.59)(0.710,0.477){7}{\rule{0.660pt}{0.115pt}}
\multiput(1008.00,730.17)(5.630,5.000){2}{\rule{0.330pt}{0.400pt}}
\multiput(1015.00,736.61)(1.579,0.447){3}{\rule{1.167pt}{0.108pt}}
\multiput(1015.00,735.17)(5.579,3.000){2}{\rule{0.583pt}{0.400pt}}
\put(1023,737.17){\rule{1.500pt}{0.400pt}}
\multiput(1023.00,738.17)(3.887,-2.000){2}{\rule{0.750pt}{0.400pt}}
\multiput(1030.59,732.55)(0.485,-1.255){11}{\rule{0.117pt}{1.071pt}}
\multiput(1029.17,734.78)(7.000,-14.776){2}{\rule{0.400pt}{0.536pt}}
\multiput(1037.59,720.00)(0.488,0.956){13}{\rule{0.117pt}{0.850pt}}
\multiput(1036.17,720.00)(8.000,13.236){2}{\rule{0.400pt}{0.425pt}}
\multiput(1045.59,732.45)(0.485,-0.645){11}{\rule{0.117pt}{0.614pt}}
\multiput(1044.17,733.73)(7.000,-7.725){2}{\rule{0.400pt}{0.307pt}}
\multiput(1052.00,726.59)(0.492,0.485){11}{\rule{0.500pt}{0.117pt}}
\multiput(1052.00,725.17)(5.962,7.000){2}{\rule{0.250pt}{0.400pt}}
\multiput(1059.00,731.93)(0.569,-0.485){11}{\rule{0.557pt}{0.117pt}}
\multiput(1059.00,732.17)(6.844,-7.000){2}{\rule{0.279pt}{0.400pt}}
\multiput(1067.00,726.61)(1.355,0.447){3}{\rule{1.033pt}{0.108pt}}
\multiput(1067.00,725.17)(4.855,3.000){2}{\rule{0.517pt}{0.400pt}}
\multiput(1074.00,727.93)(0.710,-0.477){7}{\rule{0.660pt}{0.115pt}}
\multiput(1074.00,728.17)(5.630,-5.000){2}{\rule{0.330pt}{0.400pt}}
\multiput(1081.00,722.95)(1.579,-0.447){3}{\rule{1.167pt}{0.108pt}}
\multiput(1081.00,723.17)(5.579,-3.000){2}{\rule{0.583pt}{0.400pt}}
\multiput(1089.00,721.59)(0.710,0.477){7}{\rule{0.660pt}{0.115pt}}
\multiput(1089.00,720.17)(5.630,5.000){2}{\rule{0.330pt}{0.400pt}}
\put(911.0,744.0){\rule[-0.200pt]{1.927pt}{0.400pt}}
\put(1103,724.17){\rule{1.700pt}{0.400pt}}
\multiput(1103.00,725.17)(4.472,-2.000){2}{\rule{0.850pt}{0.400pt}}
\put(1096.0,726.0){\rule[-0.200pt]{1.686pt}{0.400pt}}
\multiput(1118.59,721.45)(0.485,-0.645){11}{\rule{0.117pt}{0.614pt}}
\multiput(1117.17,722.73)(7.000,-7.725){2}{\rule{0.400pt}{0.307pt}}
\multiput(1125.00,715.59)(0.671,0.482){9}{\rule{0.633pt}{0.116pt}}
\multiput(1125.00,714.17)(6.685,6.000){2}{\rule{0.317pt}{0.400pt}}
\put(1133,719.67){\rule{1.686pt}{0.400pt}}
\multiput(1133.00,720.17)(3.500,-1.000){2}{\rule{0.843pt}{0.400pt}}
\multiput(1140.59,715.32)(0.485,-1.332){11}{\rule{0.117pt}{1.129pt}}
\multiput(1139.17,717.66)(7.000,-15.658){2}{\rule{0.400pt}{0.564pt}}
\multiput(1147.59,702.00)(0.488,0.890){13}{\rule{0.117pt}{0.800pt}}
\multiput(1146.17,702.00)(8.000,12.340){2}{\rule{0.400pt}{0.400pt}}
\put(1155,714.67){\rule{1.686pt}{0.400pt}}
\multiput(1155.00,715.17)(3.500,-1.000){2}{\rule{0.843pt}{0.400pt}}
\multiput(1162.59,711.50)(0.485,-0.950){11}{\rule{0.117pt}{0.843pt}}
\multiput(1161.17,713.25)(7.000,-11.251){2}{\rule{0.400pt}{0.421pt}}
\multiput(1169.59,702.00)(0.488,0.692){13}{\rule{0.117pt}{0.650pt}}
\multiput(1168.17,702.00)(8.000,9.651){2}{\rule{0.400pt}{0.325pt}}
\put(1111.0,724.0){\rule[-0.200pt]{1.686pt}{0.400pt}}
\multiput(1184.00,711.93)(0.710,-0.477){7}{\rule{0.660pt}{0.115pt}}
\multiput(1184.00,712.17)(5.630,-5.000){2}{\rule{0.330pt}{0.400pt}}
\multiput(1191.00,706.95)(1.579,-0.447){3}{\rule{1.167pt}{0.108pt}}
\multiput(1191.00,707.17)(5.579,-3.000){2}{\rule{0.583pt}{0.400pt}}
\multiput(1199.00,705.61)(1.355,0.447){3}{\rule{1.033pt}{0.108pt}}
\multiput(1199.00,704.17)(4.855,3.000){2}{\rule{0.517pt}{0.400pt}}
\put(1177.0,713.0){\rule[-0.200pt]{1.686pt}{0.400pt}}
\multiput(1213.59,705.30)(0.488,-0.692){13}{\rule{0.117pt}{0.650pt}}
\multiput(1212.17,706.65)(8.000,-9.651){2}{\rule{0.400pt}{0.325pt}}
\multiput(1221.59,697.00)(0.485,0.569){11}{\rule{0.117pt}{0.557pt}}
\multiput(1220.17,697.00)(7.000,6.844){2}{\rule{0.400pt}{0.279pt}}
\multiput(1228.59,702.21)(0.485,-0.721){11}{\rule{0.117pt}{0.671pt}}
\multiput(1227.17,703.61)(7.000,-8.606){2}{\rule{0.400pt}{0.336pt}}
\put(1235,695.17){\rule{1.700pt}{0.400pt}}
\multiput(1235.00,694.17)(4.472,2.000){2}{\rule{0.850pt}{0.400pt}}
\multiput(1243.00,697.59)(0.710,0.477){7}{\rule{0.660pt}{0.115pt}}
\multiput(1243.00,696.17)(5.630,5.000){2}{\rule{0.330pt}{0.400pt}}
\put(1250,700.17){\rule{1.500pt}{0.400pt}}
\multiput(1250.00,701.17)(3.887,-2.000){2}{\rule{0.750pt}{0.400pt}}
\put(1257,698.17){\rule{1.700pt}{0.400pt}}
\multiput(1257.00,699.17)(4.472,-2.000){2}{\rule{0.850pt}{0.400pt}}
\multiput(1265.00,696.95)(1.355,-0.447){3}{\rule{1.033pt}{0.108pt}}
\multiput(1265.00,697.17)(4.855,-3.000){2}{\rule{0.517pt}{0.400pt}}
\multiput(1272.00,695.60)(0.920,0.468){5}{\rule{0.800pt}{0.113pt}}
\multiput(1272.00,694.17)(5.340,4.000){2}{\rule{0.400pt}{0.400pt}}
\put(1279,697.17){\rule{1.700pt}{0.400pt}}
\multiput(1279.00,698.17)(4.472,-2.000){2}{\rule{0.850pt}{0.400pt}}
\put(1287,695.17){\rule{1.500pt}{0.400pt}}
\multiput(1287.00,696.17)(3.887,-2.000){2}{\rule{0.750pt}{0.400pt}}
\put(1206.0,708.0){\rule[-0.200pt]{1.686pt}{0.400pt}}
\put(1301,693.17){\rule{1.700pt}{0.400pt}}
\multiput(1301.00,694.17)(4.472,-2.000){2}{\rule{0.850pt}{0.400pt}}
\multiput(1309.00,691.94)(0.920,-0.468){5}{\rule{0.800pt}{0.113pt}}
\multiput(1309.00,692.17)(5.340,-4.000){2}{\rule{0.400pt}{0.400pt}}
\put(1316,687.67){\rule{1.686pt}{0.400pt}}
\multiput(1316.00,688.17)(3.500,-1.000){2}{\rule{0.843pt}{0.400pt}}
\multiput(1323.00,688.60)(1.066,0.468){5}{\rule{0.900pt}{0.113pt}}
\multiput(1323.00,687.17)(6.132,4.000){2}{\rule{0.450pt}{0.400pt}}
\multiput(1331.00,690.94)(0.920,-0.468){5}{\rule{0.800pt}{0.113pt}}
\multiput(1331.00,691.17)(5.340,-4.000){2}{\rule{0.400pt}{0.400pt}}
\put(1338,686.67){\rule{1.686pt}{0.400pt}}
\multiput(1338.00,687.17)(3.500,-1.000){2}{\rule{0.843pt}{0.400pt}}
\put(1345,686.67){\rule{1.927pt}{0.400pt}}
\multiput(1345.00,686.17)(4.000,1.000){2}{\rule{0.964pt}{0.400pt}}
\put(1353,686.17){\rule{1.500pt}{0.400pt}}
\multiput(1353.00,687.17)(3.887,-2.000){2}{\rule{0.750pt}{0.400pt}}
\put(1294.0,695.0){\rule[-0.200pt]{1.686pt}{0.400pt}}
\end{picture}

%% file: FIGPLBN4.tex
\setlength{\unitlength}{0.240900pt}
\ifx\plotpoint\undefined\newsavebox{\plotpoint}\fi
\begin{picture}(1500,900)(0,0)
\font\gnuplot=cmr10 at 10pt
\gnuplot
\sbox{\plotpoint}{\rule[-0.200pt]{0.400pt}{0.400pt}}%
\put(120.0,123.0){\rule[-0.200pt]{4.818pt}{0.400pt}}
\put(100,123){\makebox(0,0)[r]{8.2}}
\put(1419.0,123.0){\rule[-0.200pt]{4.818pt}{0.400pt}}
\put(120.0,205.0){\rule[-0.200pt]{4.818pt}{0.400pt}}
\put(100,205){\makebox(0,0)[r]{8.4}}
\put(1419.0,205.0){\rule[-0.200pt]{4.818pt}{0.400pt}}
\put(120.0,287.0){\rule[-0.200pt]{4.818pt}{0.400pt}}
\put(100,287){\makebox(0,0)[r]{8.6}}
\put(1419.0,287.0){\rule[-0.200pt]{4.818pt}{0.400pt}}
\put(120.0,369.0){\rule[-0.200pt]{4.818pt}{0.400pt}}
\put(100,369){\makebox(0,0)[r]{8.8}}
\put(1419.0,369.0){\rule[-0.200pt]{4.818pt}{0.400pt}}
\put(120.0,451.0){\rule[-0.200pt]{4.818pt}{0.400pt}}
\put(100,451){\makebox(0,0)[r]{9}}
\put(1419.0,451.0){\rule[-0.200pt]{4.818pt}{0.400pt}}
\put(120.0,532.0){\rule[-0.200pt]{4.818pt}{0.400pt}}
\put(100,532){\makebox(0,0)[r]{9.2}}
\put(1419.0,532.0){\rule[-0.200pt]{4.818pt}{0.400pt}}
\put(120.0,614.0){\rule[-0.200pt]{4.818pt}{0.400pt}}
\put(100,614){\makebox(0,0)[r]{9.4}}
\put(1419.0,614.0){\rule[-0.200pt]{4.818pt}{0.400pt}}
\put(120.0,696.0){\rule[-0.200pt]{4.818pt}{0.400pt}}
\put(100,696){\makebox(0,0)[r]{9.6}}
\put(1419.0,696.0){\rule[-0.200pt]{4.818pt}{0.400pt}}
\put(120.0,778.0){\rule[-0.200pt]{4.818pt}{0.400pt}}
\put(100,778){\makebox(0,0)[r]{9.8}}
\put(1419.0,778.0){\rule[-0.200pt]{4.818pt}{0.400pt}}
\put(120.0,860.0){\rule[-0.200pt]{4.818pt}{0.400pt}}
\put(100,860){\makebox(0,0)[r]{10}}
\put(1419.0,860.0){\rule[-0.200pt]{4.818pt}{0.400pt}}
\put(120.0,123.0){\rule[-0.200pt]{0.400pt}{4.818pt}}
\put(120,82){\makebox(0,0){1}}
\put(120.0,840.0){\rule[-0.200pt]{0.400pt}{4.818pt}}
\put(285.0,123.0){\rule[-0.200pt]{0.400pt}{4.818pt}}
\put(285,82){\makebox(0,0){2}}
\put(285.0,840.0){\rule[-0.200pt]{0.400pt}{4.818pt}}
\put(450.0,123.0){\rule[-0.200pt]{0.400pt}{4.818pt}}
\put(450,82){\makebox(0,0){3}}
\put(450.0,840.0){\rule[-0.200pt]{0.400pt}{4.818pt}}
\put(615.0,123.0){\rule[-0.200pt]{0.400pt}{4.818pt}}
\put(615,82){\makebox(0,0){4}}
\put(615.0,840.0){\rule[-0.200pt]{0.400pt}{4.818pt}}
\put(780.0,123.0){\rule[-0.200pt]{0.400pt}{4.818pt}}
\put(780,82){\makebox(0,0){5}}
\put(780.0,840.0){\rule[-0.200pt]{0.400pt}{4.818pt}}
\put(944.0,123.0){\rule[-0.200pt]{0.400pt}{4.818pt}}
\put(944,82){\makebox(0,0){6}}
\put(944.0,840.0){\rule[-0.200pt]{0.400pt}{4.818pt}}
\put(1109.0,123.0){\rule[-0.200pt]{0.400pt}{4.818pt}}
\put(1109,82){\makebox(0,0){7}}
\put(1109.0,840.0){\rule[-0.200pt]{0.400pt}{4.818pt}}
\put(1274.0,123.0){\rule[-0.200pt]{0.400pt}{4.818pt}}
\put(1274,82){\makebox(0,0){8}}
\put(1274.0,840.0){\rule[-0.200pt]{0.400pt}{4.818pt}}
\put(1439.0,123.0){\rule[-0.200pt]{0.400pt}{4.818pt}}
\put(1439,82){\makebox(0,0){9}}
\put(1439.0,840.0){\rule[-0.200pt]{0.400pt}{4.818pt}}
\put(120.0,123.0){\rule[-0.200pt]{317.747pt}{0.400pt}}
\put(1439.0,123.0){\rule[-0.200pt]{0.400pt}{177.543pt}}
\put(120.0,860.0){\rule[-0.200pt]{317.747pt}{0.400pt}}
\put(779,21){\makebox(0,0){n(efolds)}}
\put(120.0,123.0){\rule[-0.200pt]{0.400pt}{177.543pt}}
\put(138,276){\usebox{\plotpoint}}
\multiput(149.00,274.95)(3.365,-0.447){3}{\rule{2.233pt}{0.108pt}}
\multiput(149.00,275.17)(11.365,-3.000){2}{\rule{1.117pt}{0.400pt}}
\multiput(165.00,273.61)(3.365,0.447){3}{\rule{2.233pt}{0.108pt}}
\multiput(165.00,272.17)(11.365,3.000){2}{\rule{1.117pt}{0.400pt}}
\put(138.0,276.0){\rule[-0.200pt]{2.650pt}{0.400pt}}
\put(208,274.67){\rule{2.409pt}{0.400pt}}
\multiput(208.00,275.17)(5.000,-1.000){2}{\rule{1.204pt}{0.400pt}}
\put(181.0,276.0){\rule[-0.200pt]{6.504pt}{0.400pt}}
\put(231,274.67){\rule{2.168pt}{0.400pt}}
\multiput(231.00,274.17)(4.500,1.000){2}{\rule{1.084pt}{0.400pt}}
\put(218.0,275.0){\rule[-0.200pt]{3.132pt}{0.400pt}}
\multiput(253.59,268.21)(0.489,-2.300){15}{\rule{0.118pt}{1.878pt}}
\multiput(252.17,272.10)(9.000,-36.103){2}{\rule{0.400pt}{0.939pt}}
\multiput(262.58,236.00)(0.493,1.567){23}{\rule{0.119pt}{1.331pt}}
\multiput(261.17,236.00)(13.000,37.238){2}{\rule{0.400pt}{0.665pt}}
\put(275,274.67){\rule{2.168pt}{0.400pt}}
\multiput(275.00,275.17)(4.500,-1.000){2}{\rule{1.084pt}{0.400pt}}
\put(240.0,276.0){\rule[-0.200pt]{3.132pt}{0.400pt}}
\put(294,274.67){\rule{2.168pt}{0.400pt}}
\multiput(294.00,274.17)(4.500,1.000){2}{\rule{1.084pt}{0.400pt}}
\multiput(303.00,274.94)(1.358,-0.468){5}{\rule{1.100pt}{0.113pt}}
\multiput(303.00,275.17)(7.717,-4.000){2}{\rule{0.550pt}{0.400pt}}
\multiput(313.00,272.60)(1.212,0.468){5}{\rule{1.000pt}{0.113pt}}
\multiput(313.00,271.17)(6.924,4.000){2}{\rule{0.500pt}{0.400pt}}
\put(322,274.17){\rule{2.100pt}{0.400pt}}
\multiput(322.00,275.17)(5.641,-2.000){2}{\rule{1.050pt}{0.400pt}}
\put(332,273.67){\rule{1.927pt}{0.400pt}}
\multiput(332.00,273.17)(4.000,1.000){2}{\rule{0.964pt}{0.400pt}}
\put(340,275.17){\rule{2.100pt}{0.400pt}}
\multiput(340.00,274.17)(5.641,2.000){2}{\rule{1.050pt}{0.400pt}}
\multiput(350.59,269.53)(0.488,-2.211){13}{\rule{0.117pt}{1.800pt}}
\multiput(349.17,273.26)(8.000,-30.264){2}{\rule{0.400pt}{0.900pt}}
\multiput(358.58,243.00)(0.491,1.173){17}{\rule{0.118pt}{1.020pt}}
\multiput(357.17,243.00)(10.000,20.883){2}{\rule{0.400pt}{0.510pt}}
\multiput(368.00,266.59)(0.943,0.482){9}{\rule{0.833pt}{0.116pt}}
\multiput(368.00,265.17)(9.270,6.000){2}{\rule{0.417pt}{0.400pt}}
\multiput(379.00,270.94)(1.358,-0.468){5}{\rule{1.100pt}{0.113pt}}
\multiput(379.00,271.17)(7.717,-4.000){2}{\rule{0.550pt}{0.400pt}}
\multiput(389.00,266.93)(0.762,-0.482){9}{\rule{0.700pt}{0.116pt}}
\multiput(389.00,267.17)(7.547,-6.000){2}{\rule{0.350pt}{0.400pt}}
\multiput(398.58,252.95)(0.491,-2.684){17}{\rule{0.118pt}{2.180pt}}
\multiput(397.17,257.48)(10.000,-47.475){2}{\rule{0.400pt}{1.090pt}}
\multiput(408.59,210.00)(0.489,3.698){15}{\rule{0.118pt}{2.944pt}}
\multiput(407.17,210.00)(9.000,57.889){2}{\rule{0.400pt}{1.472pt}}
\multiput(417.00,274.61)(2.025,0.447){3}{\rule{1.433pt}{0.108pt}}
\multiput(417.00,273.17)(7.025,3.000){2}{\rule{0.717pt}{0.400pt}}
\multiput(427.59,266.62)(0.489,-3.116){15}{\rule{0.118pt}{2.500pt}}
\multiput(426.17,271.81)(9.000,-48.811){2}{\rule{0.400pt}{1.250pt}}
\multiput(436.58,223.00)(0.491,2.789){17}{\rule{0.118pt}{2.260pt}}
\multiput(435.17,223.00)(10.000,49.309){2}{\rule{0.400pt}{1.130pt}}
\multiput(446.00,275.93)(0.494,-0.488){13}{\rule{0.500pt}{0.117pt}}
\multiput(446.00,276.17)(6.962,-8.000){2}{\rule{0.250pt}{0.400pt}}
\multiput(454.58,259.12)(0.491,-2.945){17}{\rule{0.118pt}{2.380pt}}
\multiput(453.17,264.06)(10.000,-52.060){2}{\rule{0.400pt}{1.190pt}}
\multiput(464.00,210.93)(0.645,-0.485){11}{\rule{0.614pt}{0.117pt}}
\multiput(464.00,211.17)(7.725,-7.000){2}{\rule{0.307pt}{0.400pt}}
\put(284.0,275.0){\rule[-0.200pt]{2.409pt}{0.400pt}}
\multiput(481.59,205.00)(0.489,1.660){15}{\rule{0.118pt}{1.389pt}}
\multiput(480.17,205.00)(9.000,26.117){2}{\rule{0.400pt}{0.694pt}}
\multiput(490.59,234.00)(0.489,2.999){15}{\rule{0.118pt}{2.411pt}}
\multiput(489.17,234.00)(9.000,46.996){2}{\rule{0.400pt}{1.206pt}}
\multiput(499.59,268.57)(0.488,-5.380){13}{\rule{0.117pt}{4.200pt}}
\multiput(498.17,277.28)(8.000,-73.283){2}{\rule{0.400pt}{2.100pt}}
\multiput(507.59,204.00)(0.489,1.543){15}{\rule{0.118pt}{1.300pt}}
\multiput(506.17,204.00)(9.000,24.302){2}{\rule{0.400pt}{0.650pt}}
\multiput(516.59,231.00)(0.488,2.343){13}{\rule{0.117pt}{1.900pt}}
\multiput(515.17,231.00)(8.000,32.056){2}{\rule{0.400pt}{0.950pt}}
\multiput(524.59,258.10)(0.489,-2.650){15}{\rule{0.118pt}{2.144pt}}
\multiput(523.17,262.55)(9.000,-41.549){2}{\rule{0.400pt}{1.072pt}}
\multiput(533.59,221.00)(0.488,1.616){13}{\rule{0.117pt}{1.350pt}}
\multiput(532.17,221.00)(8.000,22.198){2}{\rule{0.400pt}{0.675pt}}
\multiput(541.59,239.77)(0.488,-1.814){13}{\rule{0.117pt}{1.500pt}}
\multiput(540.17,242.89)(8.000,-24.887){2}{\rule{0.400pt}{0.750pt}}
\multiput(549.59,215.19)(0.489,-0.728){15}{\rule{0.118pt}{0.678pt}}
\multiput(548.17,216.59)(9.000,-11.593){2}{\rule{0.400pt}{0.339pt}}
\multiput(558.59,205.00)(0.489,3.698){15}{\rule{0.118pt}{2.944pt}}
\multiput(557.17,205.00)(9.000,57.889){2}{\rule{0.400pt}{1.472pt}}
\multiput(567.59,256.75)(0.488,-3.730){13}{\rule{0.117pt}{2.950pt}}
\multiput(566.17,262.88)(8.000,-50.877){2}{\rule{0.400pt}{1.475pt}}
\multiput(575.59,209.72)(0.488,-0.560){13}{\rule{0.117pt}{0.550pt}}
\multiput(574.17,210.86)(8.000,-7.858){2}{\rule{0.400pt}{0.275pt}}
\multiput(583.59,203.00)(0.489,1.310){15}{\rule{0.118pt}{1.122pt}}
\multiput(582.17,203.00)(9.000,20.671){2}{\rule{0.400pt}{0.561pt}}
\multiput(592.59,220.19)(0.488,-1.682){13}{\rule{0.117pt}{1.400pt}}
\multiput(591.17,223.09)(8.000,-23.094){2}{\rule{0.400pt}{0.700pt}}
\multiput(600.59,200.00)(0.489,13.018){15}{\rule{0.118pt}{10.056pt}}
\multiput(599.17,200.00)(9.000,203.129){2}{\rule{0.400pt}{5.028pt}}
\multiput(609.59,375.43)(0.488,-15.286){13}{\rule{0.117pt}{11.700pt}}
\multiput(608.17,399.72)(8.000,-207.716){2}{\rule{0.400pt}{5.850pt}}
\multiput(617.59,192.00)(0.489,1.718){15}{\rule{0.118pt}{1.433pt}}
\multiput(616.17,192.00)(9.000,27.025){2}{\rule{0.400pt}{0.717pt}}
\multiput(626.59,215.15)(0.488,-2.013){13}{\rule{0.117pt}{1.650pt}}
\multiput(625.17,218.58)(8.000,-27.575){2}{\rule{0.400pt}{0.825pt}}
\multiput(634.59,191.00)(0.488,4.720){13}{\rule{0.117pt}{3.700pt}}
\multiput(633.17,191.00)(8.000,64.320){2}{\rule{0.400pt}{1.850pt}}
\multiput(642.59,251.33)(0.489,-3.524){15}{\rule{0.118pt}{2.811pt}}
\multiput(641.17,257.17)(9.000,-55.165){2}{\rule{0.400pt}{1.406pt}}
\put(651,200.17){\rule{1.700pt}{0.400pt}}
\multiput(651.00,201.17)(4.472,-2.000){2}{\rule{0.850pt}{0.400pt}}
\multiput(659.59,200.00)(0.488,1.154){13}{\rule{0.117pt}{1.000pt}}
\multiput(658.17,200.00)(8.000,15.924){2}{\rule{0.400pt}{0.500pt}}
\multiput(667.59,209.84)(0.489,-2.417){15}{\rule{0.118pt}{1.967pt}}
\multiput(666.17,213.92)(9.000,-37.918){2}{\rule{0.400pt}{0.983pt}}
\multiput(676.59,176.00)(0.488,1.154){13}{\rule{0.117pt}{1.000pt}}
\multiput(675.17,176.00)(8.000,15.924){2}{\rule{0.400pt}{0.500pt}}
\multiput(684.59,194.00)(0.488,11.324){13}{\rule{0.117pt}{8.700pt}}
\multiput(683.17,194.00)(8.000,153.943){2}{\rule{0.400pt}{4.350pt}}
\multiput(692.59,360.79)(0.489,-1.485){15}{\rule{0.118pt}{1.256pt}}
\multiput(691.17,363.39)(9.000,-23.394){2}{\rule{0.400pt}{0.628pt}}
\multiput(701.59,307.62)(0.488,-10.135){13}{\rule{0.117pt}{7.800pt}}
\multiput(700.17,323.81)(8.000,-137.811){2}{\rule{0.400pt}{3.900pt}}
\multiput(709.59,186.00)(0.488,7.097){13}{\rule{0.117pt}{5.500pt}}
\multiput(708.17,186.00)(8.000,96.584){2}{\rule{0.400pt}{2.750pt}}
\multiput(717.59,276.98)(0.488,-5.248){13}{\rule{0.117pt}{4.100pt}}
\multiput(716.17,285.49)(8.000,-71.490){2}{\rule{0.400pt}{2.050pt}}
\put(725,213.67){\rule{1.927pt}{0.400pt}}
\multiput(725.00,213.17)(4.000,1.000){2}{\rule{0.964pt}{0.400pt}}
\multiput(733.59,215.00)(0.489,3.174){15}{\rule{0.118pt}{2.544pt}}
\multiput(732.17,215.00)(9.000,49.719){2}{\rule{0.400pt}{1.272pt}}
\multiput(742.59,250.90)(0.488,-5.909){13}{\rule{0.117pt}{4.600pt}}
\multiput(741.17,260.45)(8.000,-80.452){2}{\rule{0.400pt}{2.300pt}}
\put(473.0,205.0){\rule[-0.200pt]{1.927pt}{0.400pt}}
\multiput(758.59,180.00)(0.489,0.728){15}{\rule{0.118pt}{0.678pt}}
\multiput(757.17,180.00)(9.000,11.593){2}{\rule{0.400pt}{0.339pt}}
\multiput(767.59,193.00)(0.488,2.343){13}{\rule{0.117pt}{1.900pt}}
\multiput(766.17,193.00)(8.000,32.056){2}{\rule{0.400pt}{0.950pt}}
\multiput(775.59,229.00)(0.488,8.748){13}{\rule{0.117pt}{6.750pt}}
\multiput(774.17,229.00)(8.000,118.990){2}{\rule{0.400pt}{3.375pt}}
\multiput(783.59,328.38)(0.489,-10.455){15}{\rule{0.118pt}{8.100pt}}
\multiput(782.17,345.19)(9.000,-163.188){2}{\rule{0.400pt}{4.050pt}}
\multiput(792.59,182.00)(0.488,1.616){13}{\rule{0.117pt}{1.350pt}}
\multiput(791.17,182.00)(8.000,22.198){2}{\rule{0.400pt}{0.675pt}}
\multiput(800.00,205.93)(0.671,-0.482){9}{\rule{0.633pt}{0.116pt}}
\multiput(800.00,206.17)(6.685,-6.000){2}{\rule{0.317pt}{0.400pt}}
\multiput(808.59,201.00)(0.488,1.220){13}{\rule{0.117pt}{1.050pt}}
\multiput(807.17,201.00)(8.000,16.821){2}{\rule{0.400pt}{0.525pt}}
\multiput(816.59,220.00)(0.489,0.902){15}{\rule{0.118pt}{0.811pt}}
\multiput(815.17,220.00)(9.000,14.316){2}{\rule{0.400pt}{0.406pt}}
\multiput(825.59,233.72)(0.488,-0.560){13}{\rule{0.117pt}{0.550pt}}
\multiput(824.17,234.86)(8.000,-7.858){2}{\rule{0.400pt}{0.275pt}}
\multiput(833.59,219.94)(0.488,-2.079){13}{\rule{0.117pt}{1.700pt}}
\multiput(832.17,223.47)(8.000,-28.472){2}{\rule{0.400pt}{0.850pt}}
\multiput(841.59,195.00)(0.488,5.843){13}{\rule{0.117pt}{4.550pt}}
\multiput(840.17,195.00)(8.000,79.556){2}{\rule{0.400pt}{2.275pt}}
\multiput(849.59,284.00)(0.488,4.852){13}{\rule{0.117pt}{3.800pt}}
\multiput(848.17,284.00)(8.000,66.113){2}{\rule{0.400pt}{1.900pt}}
\multiput(857.59,335.81)(0.489,-6.844){15}{\rule{0.118pt}{5.344pt}}
\multiput(856.17,346.91)(9.000,-106.907){2}{\rule{0.400pt}{2.672pt}}
\multiput(866.59,240.00)(0.488,2.013){13}{\rule{0.117pt}{1.650pt}}
\multiput(865.17,240.00)(8.000,27.575){2}{\rule{0.400pt}{0.825pt}}
\multiput(874.59,263.74)(0.488,-2.145){13}{\rule{0.117pt}{1.750pt}}
\multiput(873.17,267.37)(8.000,-29.368){2}{\rule{0.400pt}{0.875pt}}
\multiput(882.59,238.00)(0.489,0.728){15}{\rule{0.118pt}{0.678pt}}
\multiput(881.17,238.00)(9.000,11.593){2}{\rule{0.400pt}{0.339pt}}
\multiput(891.59,245.60)(0.488,-1.550){13}{\rule{0.117pt}{1.300pt}}
\multiput(890.17,248.30)(8.000,-21.302){2}{\rule{0.400pt}{0.650pt}}
\multiput(899.59,220.57)(0.488,-1.880){13}{\rule{0.117pt}{1.550pt}}
\multiput(898.17,223.78)(8.000,-25.783){2}{\rule{0.400pt}{0.775pt}}
\multiput(907.59,198.00)(0.488,16.078){13}{\rule{0.117pt}{12.300pt}}
\multiput(906.17,198.00)(8.000,218.471){2}{\rule{0.400pt}{6.150pt}}
\multiput(915.59,395.28)(0.489,-14.591){15}{\rule{0.118pt}{11.256pt}}
\multiput(914.17,418.64)(9.000,-227.639){2}{\rule{0.400pt}{5.628pt}}
\multiput(924.59,191.00)(0.488,1.550){13}{\rule{0.117pt}{1.300pt}}
\multiput(923.17,191.00)(8.000,21.302){2}{\rule{0.400pt}{0.650pt}}
\multiput(932.59,215.00)(0.488,14.229){13}{\rule{0.117pt}{10.900pt}}
\multiput(931.17,215.00)(8.000,193.376){2}{\rule{0.400pt}{5.450pt}}
\multiput(940.59,386.38)(0.488,-14.031){13}{\rule{0.117pt}{10.750pt}}
\multiput(939.17,408.69)(8.000,-190.688){2}{\rule{0.400pt}{5.375pt}}
\multiput(948.59,218.00)(0.489,2.126){15}{\rule{0.118pt}{1.744pt}}
\multiput(947.17,218.00)(9.000,33.379){2}{\rule{0.400pt}{0.872pt}}
\multiput(957.59,242.34)(0.488,-3.862){13}{\rule{0.117pt}{3.050pt}}
\multiput(956.17,248.67)(8.000,-52.670){2}{\rule{0.400pt}{1.525pt}}
\multiput(965.59,196.00)(0.488,1.286){13}{\rule{0.117pt}{1.100pt}}
\multiput(964.17,196.00)(8.000,17.717){2}{\rule{0.400pt}{0.550pt}}
\multiput(973.59,216.00)(0.488,0.560){13}{\rule{0.117pt}{0.550pt}}
\multiput(972.17,216.00)(8.000,7.858){2}{\rule{0.400pt}{0.275pt}}
\multiput(981.59,225.00)(0.489,0.728){15}{\rule{0.118pt}{0.678pt}}
\multiput(980.17,225.00)(9.000,11.593){2}{\rule{0.400pt}{0.339pt}}
\multiput(990.59,229.49)(0.488,-2.541){13}{\rule{0.117pt}{2.050pt}}
\multiput(989.17,233.75)(8.000,-34.745){2}{\rule{0.400pt}{1.025pt}}
\multiput(998.00,199.59)(0.569,0.485){11}{\rule{0.557pt}{0.117pt}}
\multiput(998.00,198.17)(6.844,7.000){2}{\rule{0.279pt}{0.400pt}}
\multiput(1006.59,206.00)(0.488,2.739){13}{\rule{0.117pt}{2.200pt}}
\multiput(1005.17,206.00)(8.000,37.434){2}{\rule{0.400pt}{1.100pt}}
\multiput(1014.59,240.02)(0.489,-2.359){15}{\rule{0.118pt}{1.922pt}}
\multiput(1013.17,244.01)(9.000,-37.010){2}{\rule{0.400pt}{0.961pt}}
\multiput(1023.59,207.00)(0.488,2.211){13}{\rule{0.117pt}{1.800pt}}
\multiput(1022.17,207.00)(8.000,30.264){2}{\rule{0.400pt}{0.900pt}}
\multiput(1031.59,241.00)(0.488,2.079){13}{\rule{0.117pt}{1.700pt}}
\multiput(1030.17,241.00)(8.000,28.472){2}{\rule{0.400pt}{0.850pt}}
\multiput(1039.59,264.91)(0.488,-2.409){13}{\rule{0.117pt}{1.950pt}}
\multiput(1038.17,268.95)(8.000,-32.953){2}{\rule{0.400pt}{0.975pt}}
\multiput(1047.59,229.68)(0.489,-1.834){15}{\rule{0.118pt}{1.522pt}}
\multiput(1046.17,232.84)(9.000,-28.841){2}{\rule{0.400pt}{0.761pt}}
\multiput(1056.59,204.00)(0.488,2.937){13}{\rule{0.117pt}{2.350pt}}
\multiput(1055.17,204.00)(8.000,40.122){2}{\rule{0.400pt}{1.175pt}}
\multiput(1064.59,249.00)(0.488,1.418){13}{\rule{0.117pt}{1.200pt}}
\multiput(1063.17,249.00)(8.000,19.509){2}{\rule{0.400pt}{0.600pt}}
\put(1072,269.67){\rule{1.927pt}{0.400pt}}
\multiput(1072.00,270.17)(4.000,-1.000){2}{\rule{0.964pt}{0.400pt}}
\multiput(1080.59,270.00)(0.489,6.436){15}{\rule{0.118pt}{5.033pt}}
\multiput(1079.17,270.00)(9.000,100.553){2}{\rule{0.400pt}{2.517pt}}
\multiput(1089.59,346.34)(0.488,-10.862){13}{\rule{0.117pt}{8.350pt}}
\multiput(1088.17,363.67)(8.000,-147.669){2}{\rule{0.400pt}{4.175pt}}
\multiput(1097.59,216.00)(0.488,1.616){13}{\rule{0.117pt}{1.350pt}}
\multiput(1096.17,216.00)(8.000,22.198){2}{\rule{0.400pt}{0.675pt}}
\multiput(1105.59,235.81)(0.488,-1.484){13}{\rule{0.117pt}{1.250pt}}
\multiput(1104.17,238.41)(8.000,-20.406){2}{\rule{0.400pt}{0.625pt}}
\multiput(1113.59,218.00)(0.489,1.543){15}{\rule{0.118pt}{1.300pt}}
\multiput(1112.17,218.00)(9.000,24.302){2}{\rule{0.400pt}{0.650pt}}
\multiput(1122.59,245.00)(0.488,0.560){13}{\rule{0.117pt}{0.550pt}}
\multiput(1121.17,245.00)(8.000,7.858){2}{\rule{0.400pt}{0.275pt}}
\put(1130,252.67){\rule{1.927pt}{0.400pt}}
\multiput(1130.00,253.17)(4.000,-1.000){2}{\rule{0.964pt}{0.400pt}}
\multiput(1138.59,248.23)(0.488,-1.352){13}{\rule{0.117pt}{1.150pt}}
\multiput(1137.17,250.61)(8.000,-18.613){2}{\rule{0.400pt}{0.575pt}}
\multiput(1146.59,228.08)(0.489,-1.077){15}{\rule{0.118pt}{0.944pt}}
\multiput(1145.17,230.04)(9.000,-17.040){2}{\rule{0.400pt}{0.472pt}}
\multiput(1155.59,213.00)(0.488,10.928){13}{\rule{0.117pt}{8.400pt}}
\multiput(1154.17,213.00)(8.000,148.565){2}{\rule{0.400pt}{4.200pt}}
\multiput(1163.59,379.00)(0.488,26.975){13}{\rule{0.117pt}{20.550pt}}
\multiput(1162.17,379.00)(8.000,366.347){2}{\rule{0.400pt}{10.275pt}}
\multiput(1171.59,675.51)(0.488,-35.625){13}{\rule{0.117pt}{27.100pt}}
\multiput(1170.17,731.75)(8.000,-483.753){2}{\rule{0.400pt}{13.550pt}}
\multiput(1179.59,245.00)(0.489,-0.786){15}{\rule{0.118pt}{0.722pt}}
\multiput(1178.17,246.50)(9.000,-12.501){2}{\rule{0.400pt}{0.361pt}}
\multiput(1188.59,234.00)(0.488,0.956){13}{\rule{0.117pt}{0.850pt}}
\multiput(1187.17,234.00)(8.000,13.236){2}{\rule{0.400pt}{0.425pt}}
\multiput(1196.59,249.00)(0.488,2.277){13}{\rule{0.117pt}{1.850pt}}
\multiput(1195.17,249.00)(8.000,31.160){2}{\rule{0.400pt}{0.925pt}}
\multiput(1204.59,270.92)(0.488,-3.994){13}{\rule{0.117pt}{3.150pt}}
\multiput(1203.17,277.46)(8.000,-54.462){2}{\rule{0.400pt}{1.575pt}}
\multiput(1212.59,223.00)(0.489,34.396){15}{\rule{0.118pt}{26.367pt}}
\multiput(1211.17,223.00)(9.000,536.275){2}{\rule{0.400pt}{13.183pt}}
\multiput(1221.59,694.03)(0.488,-38.003){13}{\rule{0.117pt}{28.900pt}}
\multiput(1220.17,754.02)(8.000,-516.017){2}{\rule{0.400pt}{14.450pt}}
\multiput(1229.59,233.85)(0.488,-1.154){13}{\rule{0.117pt}{1.000pt}}
\multiput(1228.17,235.92)(8.000,-15.924){2}{\rule{0.400pt}{0.500pt}}
\multiput(1237.59,220.00)(0.488,1.352){13}{\rule{0.117pt}{1.150pt}}
\multiput(1236.17,220.00)(8.000,18.613){2}{\rule{0.400pt}{0.575pt}}
\multiput(1245.00,241.59)(0.933,0.477){7}{\rule{0.820pt}{0.115pt}}
\multiput(1245.00,240.17)(7.298,5.000){2}{\rule{0.410pt}{0.400pt}}
\multiput(1254.59,238.74)(0.488,-2.145){13}{\rule{0.117pt}{1.750pt}}
\multiput(1253.17,242.37)(8.000,-29.368){2}{\rule{0.400pt}{0.875pt}}
\multiput(1262.59,213.00)(0.488,11.588){13}{\rule{0.117pt}{8.900pt}}
\multiput(1261.17,213.00)(8.000,157.528){2}{\rule{0.400pt}{4.450pt}}
\multiput(1270.59,386.72)(0.488,-0.560){13}{\rule{0.117pt}{0.550pt}}
\multiput(1269.17,387.86)(8.000,-7.858){2}{\rule{0.400pt}{0.275pt}}
\multiput(1278.59,373.57)(0.488,-1.880){13}{\rule{0.117pt}{1.550pt}}
\multiput(1277.17,376.78)(8.000,-25.783){2}{\rule{0.400pt}{0.775pt}}
\multiput(1286.59,331.03)(0.489,-6.145){15}{\rule{0.118pt}{4.811pt}}
\multiput(1285.17,341.01)(9.000,-96.014){2}{\rule{0.400pt}{2.406pt}}
\multiput(1295.00,243.94)(1.066,-0.468){5}{\rule{0.900pt}{0.113pt}}
\multiput(1295.00,244.17)(6.132,-4.000){2}{\rule{0.450pt}{0.400pt}}
\multiput(1303.00,239.93)(0.494,-0.488){13}{\rule{0.500pt}{0.117pt}}
\multiput(1303.00,240.17)(6.962,-8.000){2}{\rule{0.250pt}{0.400pt}}
\multiput(1311.59,233.00)(0.488,10.928){13}{\rule{0.117pt}{8.400pt}}
\multiput(1310.17,233.00)(8.000,148.565){2}{\rule{0.400pt}{4.200pt}}
\multiput(1319.59,370.17)(0.489,-8.941){15}{\rule{0.118pt}{6.944pt}}
\multiput(1318.17,384.59)(9.000,-139.586){2}{\rule{0.400pt}{3.472pt}}
\multiput(1328.59,239.19)(0.488,-1.682){13}{\rule{0.117pt}{1.400pt}}
\multiput(1327.17,242.09)(8.000,-23.094){2}{\rule{0.400pt}{0.700pt}}
\multiput(1336.59,219.00)(0.488,1.286){13}{\rule{0.117pt}{1.100pt}}
\multiput(1335.17,219.00)(8.000,17.717){2}{\rule{0.400pt}{0.550pt}}
\multiput(1344.59,234.02)(0.488,-1.418){13}{\rule{0.117pt}{1.200pt}}
\multiput(1343.17,236.51)(8.000,-19.509){2}{\rule{0.400pt}{0.600pt}}
\multiput(1352.59,217.00)(0.489,3.815){15}{\rule{0.118pt}{3.033pt}}
\multiput(1351.17,217.00)(9.000,59.704){2}{\rule{0.400pt}{1.517pt}}
\multiput(1361.59,278.02)(0.488,-1.418){13}{\rule{0.117pt}{1.200pt}}
\multiput(1360.17,280.51)(8.000,-19.509){2}{\rule{0.400pt}{0.600pt}}
\put(750.0,180.0){\rule[-0.200pt]{1.927pt}{0.400pt}}
\end{picture}

%% file: new0007043.bbl
\clearpage
\newpage 
\begin{thebibliography}{8}
\bibitem{TRA}J. H. Traschen and R. H. Brandenberger, Phys. Rev. 
D42 (1990) 2491 ;Y. Shtanov, J. H. Traschen and R. H. 
Brandenberger, Phys. Rev. D51 (1995) 5438. 
\bibitem{KLS1}L. A. Kofman, A. Linde and A. A. Starobinsky, 
Phys. Rev. Lett. 73 (1994) 3195.
\bibitem{etc}D. Boyanovsky, H. J. de Vega, R. Holman, D.-S. Lee and
 A. Singh Phys. Rev. D51 (1995) 4419;M. Yoshimura, Prog. Theor.
 Phys. 94 (1995) 8873;D. Kaiser, Phys. Rev. D53 (1996) 1776;
 H. Fujisaki, K. Kumekawa, M. Yamaguchi and M. Yoshimura,
 Phys. Rev. D{\bf 53}, 6805 (1996); J. Baacke, K. Heitmann and C. Patzold, 
 Phys. Rev. D{\bf 56}, 6556 (1997).
\bibitem{KLS2}L. A. Kofman, A. Linde and A. A. Starobinsky, 
Phys. Rev. D {\bf 56}, 3258, (1997).
\bibitem{K} L.A.Kofman,In:{\it Relativistic Astrophysics: A Conference 
 in Honor of Igor Novikof's 60th Birthday.} Copenhagen 1996,Eds.
 B. Jones and D. Markovitch, Cambridge University Press; also at
 astro-ph/9605155.
\bibitem{KT} S. Yu. Khlebnikov and I. I. Tkachev, Phys. Rev. Lett. 
{\bf 77}, 219 (1996); Phys. Lett. B {\bf 390}, 80 (1997);
 Phys. Rev. Lett. {\bf 79}, 1607 (1997).
\bibitem{Boy}D. Boyanovsky, D. Cormier, H. J. de Vega, R. Holman and 
S. P. Kumar, Proc. VIth Erice Chalonge School on Astrofundamental Physics,
Ed.N. S. Vanchez and A. Zichichi (Kluwer,1998) [hep-ph/9801453] and
references therein.
\bibitem{Fin} F. Finelli and R. Brandenberger, 
Phys. Rev. Lett. 82, (1999) 1362 , hep-ph/9809490.
\bibitem{PS}D. Polarski and A. A. 
Starobinsky, Class. Quant. Grav., 13 (1996) 377.
\bibitem{BST} J. M. Bardeen,P. J. Steinhardt and M. S. Turner,
Phys. Rev. D 28 (1983) 679,.
\bibitem{LYTH}D. H. Lyth, Phys. Rev. D31, (1985) 1792.
 \bibitem{MUK}V. F. Mukhanov, H. A. Feldman and R. H. Brandenburger, 
Phys. Rep. 215 (1992) 203.
\bibitem{LLbook} A. R. Liddle and D. H. Lyth, 'Cosmological Inflation and
 Large Scale Structure' (Camb.Univ.Press., 2000)
\bibitem{LL}A. R. Liddle and D. H. Lyth, Phys. Rep. 231 (1993) 1.
\bibitem{WMLL}D. Wands, K. A. Malik , D. H. Lyth and A. R. Liddle, 
 astro-ph 0003278.
\bibitem{LLMW}A. R. Liddle. D. H. Lyth, K. A. Malik and D. Wands, 
 Phys. Rev. D61 (2000) 103508; hep-ph/9912473.
\bibitem{BASS} B. A. Bassett, D. I. Kaiser and R. Maartens, Phys. Lett.
 B455 (1999),84:hep-ph/9808404;
B. A. Bassett, D. I. Kaiser, F. Tamburini and R. Maartens, 
 Nucl. Phys.B 561 (1999),188; hep-ph/9901319.
\bibitem{BGMK}B. A. Bassett, C. Gordon, R. Maartens and D. I. Kaiser, 
 Phys.Rev. D61 (2000), 061302; hep-ph/9909482. 
\bibitem{HMoor}A. B. Henriques and R. G. Moorhouse, Phys. Rev. D,
to be published, hep-ph/0003141.
\bibitem{DER}N. Deruelle and V. F. Mukhanov, Phys. Rev. D52, 
 (1995) 5549.
\bibitem{BassV3}B. A. Bassett, D. I. Kaiser, F. Tamburini and R. Maartens, 
 Nucl. Phys.B 561 (1999),188; hep-ph/9901319v3.
\bibitem{JS} K. Jedamzik and G. Sigl, Phys. Rev.D61 (2000) 023519;
 hep-ph/9906287.
\bibitem{KKLT} S. Khlebnikov, L. Kofman, A. Linde and  I. Tkachev,
 hep-ph/9804425.

\end{thebibliography}
